\documentclass[aps,prc,preprint,superscriptaddress,nofootinbib]{revtex4}
\usepackage[pass,a4paper]{geometry}

\usepackage{graphicx}
\usepackage{booktabs}
\usepackage{hyperref}
\usepackage{amssymb,amsmath,amsfonts}
\usepackage{slashed}
\usepackage{xcolor}

\begin{document}
    
    
    \title{Neutron Star Matter as a Relativistic Fermi Liquid}

    \author{Bengt Friman}
\affiliation{GSI Helmholtzzentrum f\"{u}r Schwerionenforschung, D-64291 Darmstadt, Germany}
\affiliation{ExtreMe Matter Institute (EMMI) {\it at} GSI, D-64291 Darmstadt, Germany}
\author{Wolfram Weise}
\affiliation{Physics Department, Technical University of Munich, D-85748 Garching, Germany}
\affiliation{ExtreMe Matter Institute (EMMI) {\it at} GSI, D-64291 Darmstadt, Germany}
\affiliation{Department of Physics, The University of Tokyo, 7-3-1 Hongo, Bunkyo-ku, Tokyo 113-0033, Japan}
\date{\today}

\vspace{5cm}

\begin{abstract}
    The equation of state (EoS) of neutron star matter, constrained by the existence of two-solar-mass stars and gravitational wave signals from neutron star mergers, is analysed using the Landau theory of relativistic Fermi liquids. While the phase diagram of dense and cold QCD matter is still open for scenarios ranging from hadronic to quark matter in the center of neutron stars, a Fermi-liquid treatment is motivated by a microscopic approach starting from a chiral nucleon-meson field theory combined with nonperturbative functional renormalization group methods. In this scheme effects of multipionic fluctuations and repulsive nuclear many-body correlations suggest that the transition to chiral symmetry restoration is shifted to densities above those typically encountered in the neutron star core. Under such conditions a Fermi-liquid description in terms of nucleon quasiparticles appears to be justified. The leading Landau parameters are derived and discussed. Our results are contrasted with a well-known Fermi liquid, namely liquid $^3$He.
\end{abstract}

\maketitle

\section{Introduction}

The physics of neutron stars entered a new era \cite{Watts2016,Ozel2016} with the observation of massive (two-solar-mass) pulsars \cite{Demorest2010, Fonseca2016, Antoniadis2013}, which set stringent constraints on the stiffness of the equation of state (EoS) of cold and dense baryonic matter, further stressed by the recent Shapiro delay measurement \cite{Cromartie2019} of a neutron star with mass $M \approx 2.17 M_\odot$. The detection of gravitational wave signals from two merging neutron stars \cite{Abbott2017} has furnished additional important information on the EoS, by providing limits for the tidal deformability and neutron star radii \cite{Most2018,De2018}. 

The composition and properties of strongly interacting matter at high densities and temperatures is a topic of continuing interest. Lattice QCD provides information on the properties of QCD matter at high temperatures and vanishing or small baryon chemical potential. However, owing to the sign problem, lattice QCD cannot at present reliably address the physics of highly compressed cold matter as it is realized in the central region of neutron stars. In this context, a broad variety of options are under discussion. A viable hypothesis invokes the description of matter in the core of neutron stars in terms of hadronic degrees of freedom (baryons and mesons) with strong many-body correlations \cite{APR1998}. Alternative descriptions involve a (possibly smooth) transition from hadronic matter to some form of quark matter \cite{Baym2018, McLerran2018}. The former option suggests that neutron star matter can be viewed as a relativistic Fermi liquid \cite{Landau1,Landau2,Landau3}, composed of neutron quasiparticles plus a small fraction of protons. This is the basic theme of the present study, in which we explore how neutron star matter behaves as a strongly coupled fermionic many-body system and how it compares with textbook Fermi systems \cite{Abrikosov1959,BP1991}, such as liquid $^3$He. Relativistic Landau Fermi-liquid theory \cite{BC1976} is an appropriate framework to address such questions. 
 
In the present study we are guided by a microscopic equation of state of neutron star matter derived from a chiral nucleon-meson (ChNM) field theory within the nonperturbative functional renormalization group (FRG) framework \cite{DW2015,DW2017}. The low-density neutron star crust region is parametrized by the Skyrme-Lyon (SLy) EoS \cite{DH2001}. The ChNM-FRG equation of state takes over at baryon densities $\rho > 0.3~\rho_0$ (with $\rho_0 = 0.16$ fm$^{-3}$, the equilibrium density of normal nuclear matter) and extends into the inner core region of the neutron star. Beta equilibrium conditions with both electrons and muons are routinely incorporated and imply a proton fraction of about 5\% in the core. The resulting EoS satisfies all empirical constraints from astrophysical observations. It is also constructed such that properties of symmetric and asymmetric nuclear matter are consistent with nuclear physics constraints (e.g., the binding energy and equilibrium density of $N=Z$ nuclear matter, the compressibility modulus, the critical temperature and density of the liquid-gas phase transition, etc.) \cite{DW2017}.

For neutron star matter, the presence of hyperons (in particular $\Lambda$ hyperons) would soften the equation of state such that a maximum mass around $2\,M_\odot$ could not possibly be reached~\cite{Djapo:2008au,Lonardoni:2014bwa}. Several recent and ongoing investigations~\cite{Lonardoni:2014bwa,Haidenbauer:2016vfq} point out that repulsive hyperon-nuclear three-body forces may be capable of preventing the appearance of hyperons in neutron stars alltogether and thus maintain the necessary stiffness of the EoS. 

Given its nonperturbative nature, the ChNM-FRG equation of state is in principle not limited to low densities, in contrast to the EoS derived from (perturbative) chiral effective field theory (ChEFT). The two schemes are mutually consistent in the low-density range, $\rho\lesssim 2\,\rho_0$. The window of applicability for the ChNM-FRG model potentially extends considerably beyond these densities. Its limitation is ultimately determined by the transition from the (spontaneously broken) Nambu-Goldstone phase to the (restored) Wigner-Weyl realization of chiral symmetry. In the presence of multipion fluctuations and nuclear many-body correlations, all treated nonperturbatively by solving the FRG equations, it turns out that this transition from spontaneously broken to restored chiral symmetry is, in this model, shifted to baryon densities well above five times $\rho_0$. For comparison, the central density of a $2\,M_\odot$ neutron star, computed with such a model, does not exceed about $5\,\rho_0$. In such an approach the neutron star core is thus composed entirely of ``nonexotic" (nucleonic and pionic) degrees of freedom in the presence of strongly repulsive correlations. 

The input for the effective action of the ChNM-FRG model \cite{DW2015,DW2017} is prepared such that it is consistent with well-known ground state properties of nuclear matter, nuclear thermodynamics including the critical point of the liquid-gas phase transition \cite{Elliot2013}, and {\it ab initio} neutron matter computations using realistic nucleon-nucleon interactions \cite{Gandolfi2014, Heb2013, HK2017,LH2018,Dri2019}. With isospin-dependent interactions that reproduce the empirical asymmetry energy of $(32\pm 2)$ MeV at nuclear saturation density {\cite{Baldo2016}, the resulting EoS does indeed yield stable neutron stars that satisfy the $2\,M_\odot$ constraint. Moreover, it is located well within the band of equations of state,  $P({\cal E})$ [pressure as a function of energy density] that have been extracted in refs.\,\cite{Annala2018,Vuorinen2018}. It is also consistent with the EoS band deduced by the LIGO and Virgo Collaborations from the gravitational wave signals generated by the neutron star merger event GW170817 \cite{Abbott2018}, as well as with the EoS recently deduced from neutron star data using neural network techniques \cite{FFM2019}. 

The emerging picture of a neutron star with such a generic chiral EoS is that of a relativistic liquid of nucleons (primarily neutrons) correlated by strong multipion fields and short-distance repulsive forces. A description in terms of fermionic quasiparticles thus suggests a treatment using Landau Fermi liquid theory. The (small) proton fraction in neutron star matter plays a non-negligible role in quantitative astrophysical considerations. It is of minor importance, however, for our purpose of studying Fermi-liquid properties which can therefore be focused primarily on pure neutron matter. The aim of the present work is therefore to identify quasiparticle properties of the neutrons as the dominant Fermionic degrees of freedom in the microscopic chiral FRG equation of state~\cite{DW2015,DW2017} that satisfies all observational constraints, and to study and interpret the corresponding Fermi liquid properties.

In the following sections we prepare the framework for such a Fermi-liquid description. We then proceed to deduce and interpret the lowest Landau parameters of the spin-independent quasiparticle interaction. These quantify the strength of the correlations and characterize the bulk properties of matter in the deep interior of neutron stars.

\section{Landau theory of relativistic Fermi liquids \label{sec:Landau}}

\subsection{Reminder of Landau Fermi-liquid theory}

It is useful to start by first recalling the nonrelativistic theory. For nuclear many-body systems, this limit is realized at low baryon densities $\rho$ where the Fermi momentum, $p_F= (6\pi^2\rho/\nu)^{1/3}$ (with spin-isospin degeneracy $\nu = 2$ for neutron matter and $\nu = 4$ for symmetric nuclear matter), is small compared to the quasiparticle effective mass. 

Most of our discussion will be restricted to vanishing temperatures, i.e., $T=0$. Superfluidity is ignored since the EoS in the density range of interest is practically unaffected by pairing. In Landau's theory of normal Fermi liquids \cite{Landau1,Landau2,Landau3,Abrikosov1959,BP1991}, the variation of the energy of the system with changes of the quasiparticle occupation numbers is given by
\begin{equation}\label{eq:Landau-energy}
\delta E = \sum_p \varepsilon_p\,\delta n_p + \frac{1}{2\,V}\,\sum_{p,p^\prime} {\cal F}_{pp^\prime}\,\delta n_p\,\delta n_{p^\prime}~.
\end{equation}
Here $V$ is the volume,  $\varepsilon_p$ is the quasiparticle energy, ${\cal F}_{pp^\prime}$ is the quasiparticle interaction and $\delta n_p=n_p-n^{(0)}_p$ is the deviation of the quasiparticle distribution function from the ground state distribution
\begin{equation}
n^{(0)}_p=\left\{\begin{array}{cc}1,&~~~\varepsilon_p<\mu~\\
0,&~~~\varepsilon_p>\mu~,\end{array}\right.
\end{equation}
where $\mu$ is the chemical potential, or equivalently the energy of a quasiparticle on the Fermi surface. Consequently, in the ground state of a uniform system, the distribution function equals unity for quasiparticle momenta below the Fermi momentum $p_F$ and vanishes for momenta above $p_F$. The energy of a quasiparticle with momentum $p$ is given by the first variation of the energy with respect to the occupation number $n_p$
\begin{equation}
\varepsilon_p=\frac{\delta E}{\delta n_p}\,,
\end{equation}
while the quasiparticle interaction is determined by the second variation
\begin{equation}\label{eq:qp-interaction}
{\cal F}_{pp^\prime}=V\frac{\delta^2 E}{\delta n_p\,\delta n_{p\prime}}=V\frac{\delta\varepsilon_p}{\delta n_{p^\prime}}\,.
\end{equation}
For the low-lying excitations of interest in Fermi-liquid theory, the relevant quasiparticle states are near the Fermi surface. Hence, in the quasiparticle interaction one in general can set $|\boldsymbol{p}|=|\boldsymbol{p}^\prime|=p_F$.
The velocity of a quasiparticle on the Fermi surface is given by
\begin{equation}
v_F=\left(\frac{\partial \varepsilon_p}{\partial p}\right)_{p=p_F},
\end{equation}
and defines the quasiparticle effective mass $m^*$ through 
\begin{equation}
v_F=\frac{p_F}{m^*}\,.
\end{equation} 
Thus, near the Fermi surface, the quasiparticle energy takes the form
\begin{equation}
\varepsilon_p=\mu+v_F(p-p_F)\,.
\end{equation}
The density of quasiparticle states at the Fermi surface is given by
\begin{equation}\label{eq:density-of-states}
N(0)=\frac{1}{V}\sum_p\delta(\varepsilon_p-\mu)\,.
\end{equation}
Replacing the sum over $p$ by an integral (including the sums over spin and isospin degrees of freedom), one finds:
\begin{equation}
N(0)=\frac{\nu\, m^*p_F}{2 \pi^2}\,.
\end{equation}
In particular, for neutron matter, $N(0)=m^*p_F/\pi^2$. 

We now focus on pure neutron matter. For simplicity we ignore noncentral forces (e.g. spin-orbit interactions). These are non-leading effects in neutron matter which contribute only in $p$- and higher partial waves. With these restrictions the spin-dependent quasiparticle interaction is of the form
\begin{equation}\label{eq:qp-int-spin}
{\cal F}_{pp^\prime}=f_{pp^\prime}+g_{pp^\prime}\,\boldsymbol{\sigma}\cdot\boldsymbol{\sigma^\prime}.
\end{equation} 
The momentum dependence of the functions $f_{pp^\prime}$ and $g_{pp^\prime}$ is expanded in Legendre polynomials according to
\begin{equation}\label{eq:qp-int-legendre}
f_{pp^\prime}=\sum_{\ell=0}^{\infty}f_\ell\,P_\ell(\cos \theta)\,,~~~g_{pp^\prime}=\sum_{\ell=0}^{\infty}g_\ell\,P_\ell(\cos \theta)~,
\end{equation}
where $\theta$ is the angle between the two momenta $\boldsymbol{p}$ and $\boldsymbol{p}^\prime$. The coefficients in this expansion are the Landau Fermi-liquid parameters. 
It is useful to define dimensionless Landau parameters:
\begin{equation}
F_\ell=N(0)\,f_\ell~,~~~G_\ell=N(0)\,g_\ell~.
\end{equation}
In the present context, as we focus on the ground state of spin-saturated neutron matter, only the spin-independent parameters $F_\ell$ are of prime interest. 

Basic properties of the Fermi liquid can be expressed in terms of the first few Landau parameters. For example, the quasiparticle effective mass, $m^*$, is given by the spin-independent Landau parameter $F_1$ which is a measure of the velocity dependence of the quasiparticle interaction:
\begin{equation}\label{eq:F1}
\frac{m^*}{M_0}=1+\frac{F_1}{3}\,,
\end{equation}
with the free (vacuum) particle mass $M_0$. The specific heat of a Fermi liquid at low temperatures is determined by the quasiparticle effective mass as follows:
\begin{equation}
c_V=\frac{m^*p_F}{3}\,T.
\end{equation}
Finally, the incompressibility of the Fermi liquid is given by
\begin{equation}
K=9\rho\,\frac{\partial^2 {\cal E}}{\partial \rho^2}=6\frac{p_F^2}{2 m^*}(1+F_0)=\frac{3\,p_F^2}{M_0}\,\,\frac{1+F_0}{1+F_1/3},
\end{equation}
where ${\cal E}=E/V$ is the energy density, and
the speed of (first) sound in the system is given by 
\begin{equation}
c_1^2=\frac{p_F^2}{3\,M_0^2}\,\,\frac{1+F_0}{1+F_1/3}.
\end{equation}
This is the nonrelativistic sound speed with $c_1 \ll 1$. 

\subsection{Relativistic Fermi liquids}

Baryonic matter at the high densities encountered in the core of neutron stars makes a relativistic treatment mandatory. For example, the Fermi momentum in neutron matter at $\rho = 5\,\rho_0$ is $p_F\approx 0.57$\,GeV, i.e., of a magnitude comparable to the effective mass. In a relativistic Fermi liquid \cite{BC1976}, the speed of sound is given by
\begin{equation}\label{eq:soundspeed}
c_1^2=\frac{\partial P}{\partial{\cal E}}=\frac{\rho}{\mu}\frac{\partial \mu}{\partial \rho},
\end{equation}
where $P$ is the pressure, ${\cal E}$ the energy density, $\mu$ the baryon chemical potential, and we have used $d P=\rho\, d\mu$ and $d{\cal E}=\mu \,d\rho$. With 
\begin{equation}
\rho =\frac{1}{V}\sum_p \theta(\mu-\varepsilon_p),
\end{equation}
one finds that 
\begin{equation}\label{eq:d-rho-d-mu}
\frac{\partial \rho}{\partial \mu}=\frac{1}{V}\sum_p\delta(\mu-\varepsilon_p)\left(1-\frac{\partial\varepsilon_p}{\partial \mu}\right).
\end{equation}
Now, at zero temperature we have 
\begin{equation}\label{eq:d-epsilon-d-mu}
\frac{\partial\varepsilon_p}{\partial \mu}=\frac{\partial\varepsilon_p}{\partial \rho}\frac{\partial\rho}{\partial \mu}.
\end{equation}
In order to compute $\partial \varepsilon_p/\partial\rho$, we introduce a variation of the quasiparticle occupation number, which is spin independent and spherically symmetric, i.e., $\delta n_p=\eta\, \delta(p-p_F)$, and satisfies 
\begin{equation}\label{eq:delta-rho}
\delta \rho=\frac{1}{V}\sum_p \delta n_p=\frac{\nu}{(2\pi)^3}\int d^3 p\,\eta\,\delta(p-p_F)=\frac{\nu\,\eta\, p_F^2}{2\pi^2}.
\end{equation}
From (\ref{eq:qp-interaction}) and (\ref{eq:qp-int-legendre}) it follows that 
\begin{equation}
\delta \varepsilon_p=\frac{1}{V}\sum_{p^\prime}{\cal F}_{pp^\prime}\,\delta n_{p^\prime} =\frac{\nu\,\eta\, p_F^2}{2\pi^2}f_0~,
\end{equation}
and consequently that
\begin{equation}\label{eq:d-epsilon-d-rho}
\frac{\partial \varepsilon_p}{\partial \rho}=f_0~.
\end{equation}
Inserting (\ref{eq:d-epsilon-d-rho}) and (\ref{eq:d-epsilon-d-mu}) in (\ref{eq:d-rho-d-mu}) and solving for $\partial\rho/\partial\mu$, one finds 
\begin{equation}
\frac{\partial\rho}{\partial \mu}=\frac{N(0)}{1+N(0)f_0}=\frac{p_F\,m^*}{\pi^2(1+F_0)}\,.
\label{eq:24}
\end{equation}
We note that the definiton of the Landau effective mass, $m^*=p_F\left(\frac{\partial \varepsilon_p}{\partial p}\right)^{-1}_{p=p_f}$ remains unchanged in the relativistic formulation, while the corresponding effective mass relation is given by~\cite{BC1976} 
\begin{equation}\label{eq:rel-eff-mass-relation}
\frac{m^*}{\mu}=1+F_1/3\,.
\end{equation} 
Finally the squared speed of sound becomes
\begin{equation}
\label{eq:sound-speed-Landau}
c_1^2=\frac{p_F^2}{3\,\mu\, m^*}(1+F_0)=\frac{p_F^2}{3\,\mu^2}\,\,\frac{1+F_0}{1+F_1/3}\,,
\end{equation}
and the incompressibility is
\begin{equation}
K=6\,\frac{p_F^2}{2 \mu}\,\frac{1+F_0}{1+F_1/3} = 9\mu\,c_1^2\,.
\end{equation}
One notes that in these expressions, the relativistic treatment simply replaces the Fermion mass $M_0$ in the nonrelativistic forms by the baryon chemical potential 
\begin{equation}
\mu = \frac{\partial{\cal E}}{\partial\rho} = \varepsilon_{p=p_F}\,.
\end{equation}

\subsection{Relativistic quasiparticles}

Central to Landau Fermi-liquid theory is the notion of quasiparticles dressed by their interactions with the surrounding many-body system.
Consider first a simplified example with fermions moving in a scalar field. This scalar field modifies the fermion mass at nonzero densities.  The quasiparticle energy is given by
\begin{equation}\label{eq:rel-qp-energy1}
\varepsilon_p=\sqrt{p^2+M^2}~,
\end{equation}
where $M(\rho)$ is a function of the scalar field and thus of the density, with $M(\rho = 0) \equiv M_0$, the vacuum mass. The variation of the energy of the system is again given by (\ref{eq:Landau-energy}). Using (\ref{eq:rel-qp-energy1}) and (\ref{eq:qp-interaction}) we find
\begin{equation}
\delta E=\sum_p \sqrt{p^2+M^2}\,\delta n_p+\frac{1}{2}\sum_{pp^\prime}\frac{\delta \sqrt{p^2+M^2}}{\delta n_{p^\prime}}\, \delta n_p\,\delta n_{p^\prime}~.
\end{equation}
Thus, in this model the quasiparticle interaction is
\begin{equation}
{\cal F}_{pp^\prime}=V\,\frac{M}{\sqrt{p^2+M^2}}\,\frac{\delta M}{\delta n_{p^\prime}}~.
\end{equation}
Using
\begin{equation}
\frac{\delta M}{\delta n_{p^\prime}}=\frac{\partial M}{\partial\rho}\,\frac{\delta\rho}{\delta n_{p^\prime}}~.
\end{equation}
together with the first equality in (\ref{eq:delta-rho}),
\begin{equation}\label{eq:rho-variation}
\frac{\delta\rho}{\delta n_{p^\prime}}=\frac{1}{V}~,
\end{equation}
one finds:
\begin{equation}\label{eq:qp-scalar-int}
{\cal F}_{pp^\prime}=\frac{M(\rho)}{\sqrt{p_F^2+M^2(\rho)}}\,\frac{\partial M}{\partial \rho}~,
\end{equation}
where we have put the quasiparticles on the Fermi surface. Since the quasiparticle interaction is independent of the angle between $\boldsymbol{p}$ and $\boldsymbol{p}^\prime$ and independent of spin, the only nonzero Fermi-liquid parameter is $f_0$. This means that, in particular, $F_1=0$ and that the effective mass at the Fermi surface is equal to the chemical potential:
\begin{equation}
m^*=\mu=\sqrt{p_F^2+M^2(\rho)}.
\end{equation} 
Hence, the dimensionless Fermi-liquid parameter is given by
\begin{equation}
F_0=\frac{p_F\,M}{\pi^2}\,\frac{\partial M}{\partial \rho}=\frac{M}{p_F}\,\frac{\partial M}{\partial p_F}
\end{equation}
and the squared speed of sound is 
\begin{equation}
c_1^2=\frac{p_F^2}{3\,(p_F^2+M^2)}\left(1+\frac{p_F\, M}{\pi^2}\frac{\partial M}{\partial \rho}\right)\,.
\end{equation}
The in-medium nucleon mass $M(\rho)$ is usually a decreasing function of density. With $\partial M/\partial \rho <0$, the Landau parameter $F_0$ stays negative. The squared speed of sound will always be $c_1^2 < 1/3$ and approach the limit of a free ultra-relativistic gas from below. As we shall see, the equation of state of neutron star matter implies instead $c_1^2>1/3$ starting from some intermediate density, reflecting increasingly strong repulsive correlations between the quasiparticles as the density increases. 

In a relativistic theory, a repulsive short-distance interaction is most naturally viewed as being mediated by a vector field.  Consider therefore quasiparticles interacting with a vector field $V^\mu$ in addition to the scalar field. The vector field is in turn assumed to be generated by the baryon four-current $j^\mu =(\rho,\boldsymbol{j})$:
\begin{equation}
V^\mu=(V_0,\boldsymbol{V})=h\,j^\mu\,.
\end{equation}
In the mean-field approximation the vector field is a linear function of the current, i.e., $h$ is a constant. Effects beyond the mean-field limit imply a nonlinear behavior that can be incorporated by a generalized {\it ansatz}:
\begin{equation}
V^\mu=h(j^2)\,j^\mu\,,
\end{equation}
with $h$ now assumed to be a function of the Lorentz invariant $j^2=j_\mu\,j^\mu$.
In the rest frame of a system in its ground state, only the zeroth component of the baryon current, $j^0=\rho$, is nonzero. The current in the frame of an observer moving relative to the system is obtained by a Lorentz transformation.

In a general frame, the quasiparticle energy in this model can be written:
\begin{equation}\label{eq:qp-energy-II}
\varepsilon_p=\sqrt{(\boldsymbol{p}-\boldsymbol{V})^2+M^2}+V_0~,
\end{equation}
where $\boldsymbol{p}-\boldsymbol{V}$ is the kinetic momentum. Thus, the velocity of a quasiparticle is given by
\begin{equation}\label{eq:qp-velocity}
\boldsymbol{v}_p=\frac{\boldsymbol{p}-\boldsymbol{V}}{\sqrt{(\boldsymbol{p}-\boldsymbol{V})^2+M^2}}~.
\end{equation}
In the rest frame of the system, (\ref{eq:qp-energy-II}) reduces to
\begin{equation}
\varepsilon_p=\sqrt{\boldsymbol{p}^2+M^2}+V_0~,
\end{equation} 
In general, a variation of the occupation number leads to nonzero spatial components of the baryon current,
\begin{equation}\label{eq:qp-current}
\boldsymbol{j}=\frac{1}{V}\sum_p\boldsymbol{v}_p\, n_p~.
\end{equation}

The quasiparticle interaction is again obtained by varying the quasiparticle energy with respect to the occupation number:
\begin{equation}
{\cal F}_{pp^\prime}=V\,\frac{\delta\,\varepsilon_p}{\delta n_{p^\prime}}~.
\end{equation}
The following discussion is similar to that of Ref.\,\cite{Matsui1981}, although our assumptions are more general.
Given the quasiparticle energy (\ref{eq:qp-energy-II}), one thus finds:
\begin{equation}\label{eq:qp-int-II}
\frac{f_{pp^\prime}}{V}=\frac{M}{\sqrt{(\boldsymbol{p}-\boldsymbol{V})^2+M^2}}\left(\frac{\delta M}{\delta n_{p^\prime}}\right)+\frac{\delta V_0}{\delta n_{p^\prime}}-\frac{\boldsymbol{p}-\boldsymbol{V}}{\sqrt{(\boldsymbol{p}-\boldsymbol{V})^2+M^2}}\left(\frac{\delta\boldsymbol{V}}{\delta n_{p^\prime}}\right)~.
\end{equation}
At this level there are only spin-independent contributions to the quasiparticle interaction. In the rest frame of the fluid, the first term in (\ref{eq:qp-int-II}) reduces to the result already found in Eq.\,(\ref{eq:qp-scalar-int}),
\begin{equation}
f_{pp^\prime}(1)=\frac{M}{\sqrt{\boldsymbol{p}^2+M^2}}\,\frac{\partial M}{\partial \rho}~.
\end{equation}
The variation of the zeroth component of the vector field yields
\begin{equation}\label{eq:variation-V0}
\frac{\delta V_0}{\delta n_{p^\prime}}=\frac{\partial V_0}{\partial\rho}\,\frac{\delta \rho}{\delta n_{p^\prime}}+ \frac{\partial V_0}{\partial\boldsymbol{j}}\,\frac{\delta \boldsymbol{j}}{\delta n_{p^\prime}}~.
\end{equation}
The second term in (\ref{eq:variation-V0}) vanishes in the rest frame of the system, since 
\begin{equation}
\frac{\partial V_0}{\partial \boldsymbol{j}}=-h^\prime(j^2)\,\boldsymbol{j}\,\rho~,
\end{equation}
and $\boldsymbol{j}=0$ in that frame. Hence, using (\ref{eq:rho-variation}), we find
\begin{equation}
\frac{\delta V_0}{\delta n_{p^\prime}}=\frac{1}{V}\,\frac{\partial V_0}{\partial\rho}~,
\end{equation}
and for the contribution of the second term to the quasiparticle interaction (\ref{eq:qp-int-II}):
\begin{equation}
f_{pp^\prime}(2)=\frac{\partial V_0}{\partial \rho}~.
\end{equation}

The detailed derivation of the last term on the right-hand side. of Eq.\,(\ref{eq:qp-int-II}) is relegated to Appendix A. Its contribution to the quasiparticle interaction is:
\begin{equation}
f_{pp^\prime}(3)=-h(j^2)\frac{\boldsymbol{p}\cdot\boldsymbol{p}^\prime}{\mu\,\sqrt{p_F^2+M^2}}~,
\label{eq:3rdterm}
\end{equation}
Now, collecting all pieces, the resulting Landau Fermi-liquid parameters are given by
\begin{eqnarray}
F_0&=&\frac{p_F\,}{\pi^2}\,\left(M(\rho)\,\frac{\partial M}{\partial\rho}+\sqrt{p_F^2+M^2(\rho)}\,\,\frac{\partial V_0}{\partial \rho}\right)\,\label{eq:xF0}~,\\
F_1&=&-3\,\frac{h(j^2)\,\rho}{\mu}=-\frac{3V_0(\rho)}{\mu}~.
\label{eq:xF1}
\end{eqnarray}
An equivalent way of deriving the result for $F_0$ proceeds directly through the derivative of the chemical potential with the assumed ansatz
\begin{equation}
\mu = \sqrt{p_F^2+M^2(\rho)} + V_0(\rho) \equiv m^*(\rho) + V_0(\rho)\,.
\label{eq:mu}
\end{equation}
Observing that
\begin{equation}
\frac{\partial\mu}{\partial\rho} = \frac{1}{ m^*}\left[\frac{\pi^2}{ p_F} + M\frac{\partial M}{\partial\rho}\right]+\frac{\partial V_0}{ \partial\rho}\,,
\end{equation}
leads to 
\begin{equation}
1+F_0(\rho) = \frac{p_F\,m^*}{ \pi^2}\left(\frac{\partial\mu}{\partial\rho}\right)\,,
\label{eq:yF0}
\end{equation} 
[see also Eq.\,(\ref{eq:24})], while the relation for $F_1$,
\begin{equation}
1+\frac{F_1(\rho)}{ 3} = 1-\frac{V_0(\rho)}{\mu} = \frac{m^*}{\mu}\,,
\label{eq:yF1}
\end{equation}
is consistent with the nonrelativistic Eq.\,(\ref{eq:F1}).

\section{Neutron Star Equation of State and Fermi-Liquid Theory}

\subsection{Chiral FRG equation of state}

The starting point is now an equation of state, pressure $P({\cal E})$ as a function of energy density ${\cal E}$, derived from a chiral field theory of nucleons and mesons (the ChNM model), based on a linear sigma model with a nonlinear effective potential. The basic Lagrangian involves the isospin-doublet field of the nucleon, $N = (p,n)^\top$, and the chiral boson field $\phi = (\sigma,\boldsymbol{\pi})$ composed of a heavy scalar $\sigma$ and the pseudoscalar Nambu-Goldstone boson $\boldsymbol{\pi}$ of spontaneously broken chiral $SU(2)_L\times SU(2)_R$ symmetry: 
\begin{eqnarray}
{\cal L} &=& \bar{N}\left[i\gamma_\mu\partial^\mu - g(\sigma + i\gamma_5\,\boldsymbol{\tau\cdot\pi})\right]N \nonumber  \\ &+& {\frac12}\left(\partial_\mu \sigma \partial^\mu \sigma + \partial_\mu \boldsymbol{\pi}\cdot\partial^\mu \boldsymbol{\pi}\right) - {\cal U}(\sigma, \boldsymbol{\pi})+ \Delta{\cal L}~.
\label{eq:ChNM}
\end{eqnarray}
The $\Delta{\cal L}$ term of this Lagrangian represents short-distance dynamics expressed in terms of isoscalar and isovector vector fields coupled to nucleons, corresponding to contact interactions in chiral effective field theory (ChEFT). The potential ${\cal U}(\sigma, \boldsymbol{\pi})$ is written as a polynomial up to fourth order in the chiral invariant, $\chi \equiv\phi^\dagger\phi = \sigma^2 + \boldsymbol{\pi}^2$, plus a symmetry breaking piece proportional to $m_\pi^2\sigma$. This potential is constructed such as to be consistent with pion-nucleon data and selected ground state properties of nuclear matter. (For details see Refs.\,\cite{DW2015,DW2017} and a brief overview in Appendix B.)

The action $S=\int d^4x\,{\cal L}$ of the ChNM Lagrangian (\ref{eq:ChNM}) serves as input for a functional renormalization group (FRG) calculation starting at a UV scale (the chiral symmetry breaking scale $4\pi f_\pi \approx 1$ GeV, with the pion decay constant $f_\pi \approx 0.09$ GeV). This is then evolved down to the full effective action at the low momentum (IR) limit using the FRG flow equations \cite{Wet1993}. The grand-canonical potential $\Omega(T,\mu)$ is constructed \cite{DW2015,DW2017}, from which the pressure $P = -\Omega$ and the energy density ${\cal E} = -P+\mu\rho$ (at $T=0$) are derived, resulting in the equation of state $P({\cal E})$. 

Figure\,\ref{fig:1} shows this equation of state, including an estimated band of uncertainties. These uncertainties arise primarily from the input parameters of the ChNM model, in particular from varying the nuclear asymmetry energy, $A_S$, in the empirical range between 30 and 34 MeV. The solid curve with $A_S = 32$ MeV serves as our prototype EoS, referred to as the chiral FRG equation of state in the following. A rough estimate of uncertainties in the pressure $P$ amounts to $\pm 15\, \%$. For practical purposes, an accurate Pad\'e fit and a table of $P({\cal E})$ are given in Appendix C.

The squared velocity of (first) sound (\ref{eq:soundspeed})
is shown in Fig.\,\ref{fig:2}. Notably, the sound speed exceeds its canonical value for a noninteracting ultrarelativistic Fermi gas $c_1^2 = 1/3$ at densities $\rho\gtrsim 4\,\rho_0$, mainly as a consequence of repulsive multi-nucleon correlations which govern the stiffness of the EoS. A similar behavior of $c_1^2$ is reported in \cite{FFM2019} and also discussed in \cite{Tews2018}.

Important input for the chiral FRG EoS is set by nuclear physics constraints at baryon densities $\rho\lesssim 2\rho_0$. The nonperturbative chiral FRG approach is designed to be consistent with perturbative chiral effective field theory calculations \cite{Heb2013, HK2017, LH2018, Dri2019} for both symmetric nuclear matter and neutron matter in this density range. In fact, whereas the chiral FRG starts from a Lagrangian based on a linear sigma model, the pion sector of the ChEFT is built on a nonlinear sigma model, where the heavy scalar $\sigma$ field has been eliminated. While the linear and nonlinear sigma models are not equivalent at any perturbative level, resummations to all orders in the nonperturbative FRG treatment of the ChNM model should yield results that match those of the perturbative ChEFT at sufficiently low momentum scales and densities. This is indeed demonstrated by explicit calculations in Refs. \cite{DW2015, DW2017}.

Figure \ref{fig:3} displays the resulting EoS, $P({\cal E})$, in comparison with the families of equations of state obtained in Refs.\,\cite{Annala2018,Vuorinen2018}. The latter interpolate between the low-density ChEFT EoS and the high-density EoS of perturbative QCD and satisfy the astrophysical constraints from neutron star masses and the tidal deformability deduced from the recently observed neutron star merger GW170817. Clearly the ChNM-FRG EoS fits well into the allowed band.

\begin{figure*}[t]
\begin{center}
\includegraphics[height=70mm,angle=-00]{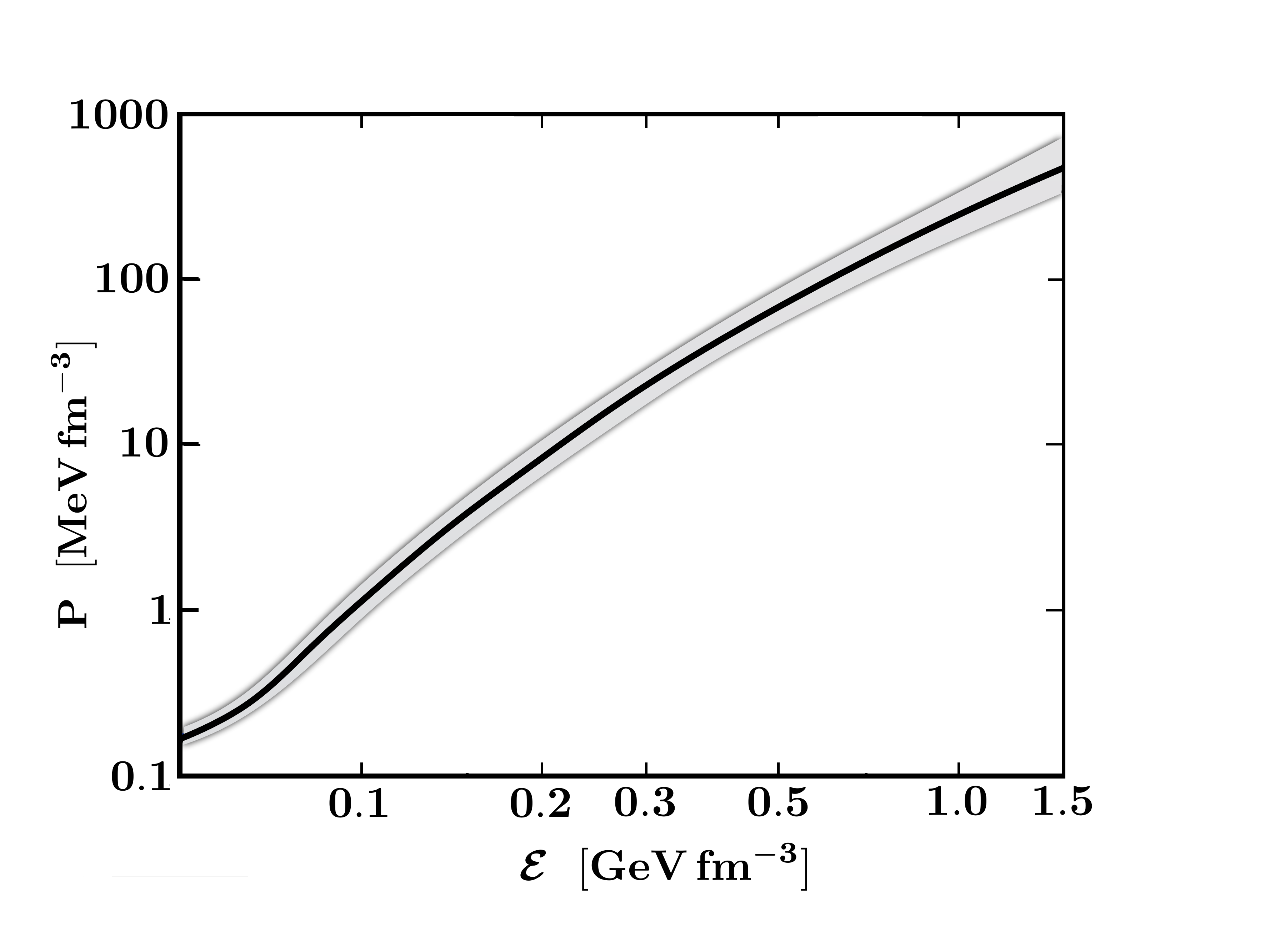}
\caption{The equation of state of neutron star matter in beta equilibrium (pressure $P$ as a function of energy density ${\cal E}$) derived from chiral nucleon-meson field theory combined with functional renormalization group equations \cite{DW2015,DW2017}. The shaded band gives an uncertainty estimate when varying the symmetry energy in the range $30 - 34$\,MeV. In the limit of very low densities ($\rho < 0.3\,\rho_0$ corresponding to ${\cal E} < 0.045$\,GeV/fm$^3$), the EoS is matched to the Skyrme-Lyon (SLy) parametrization \cite{DH2001}.}
\label{fig:1}
\end{center}
\end{figure*}

\begin{figure*}[t]
\begin{center}
\includegraphics[height=60mm,angle=-00]{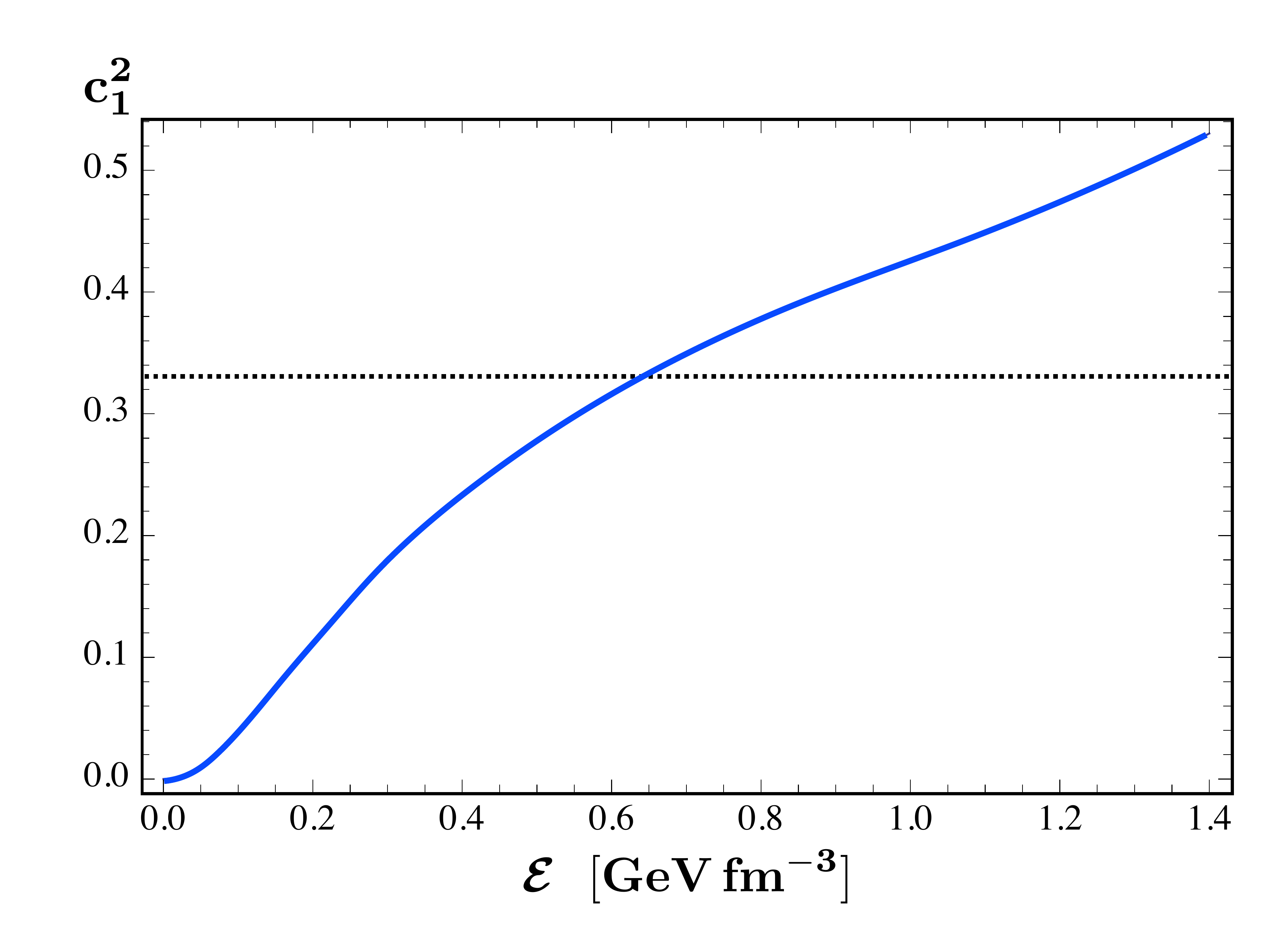}
\caption{The squared speed of sound, $c_1^2 = \partial P({\cal E})/\partial{\cal E}$, derived from the neutron star matter equation of state, Fig.\,\ref{fig:1}. }
\label{fig:2}
\end{center}
\end{figure*}

\begin{figure*}[t]
\begin{center}
\includegraphics[height=75mm,angle=-00]{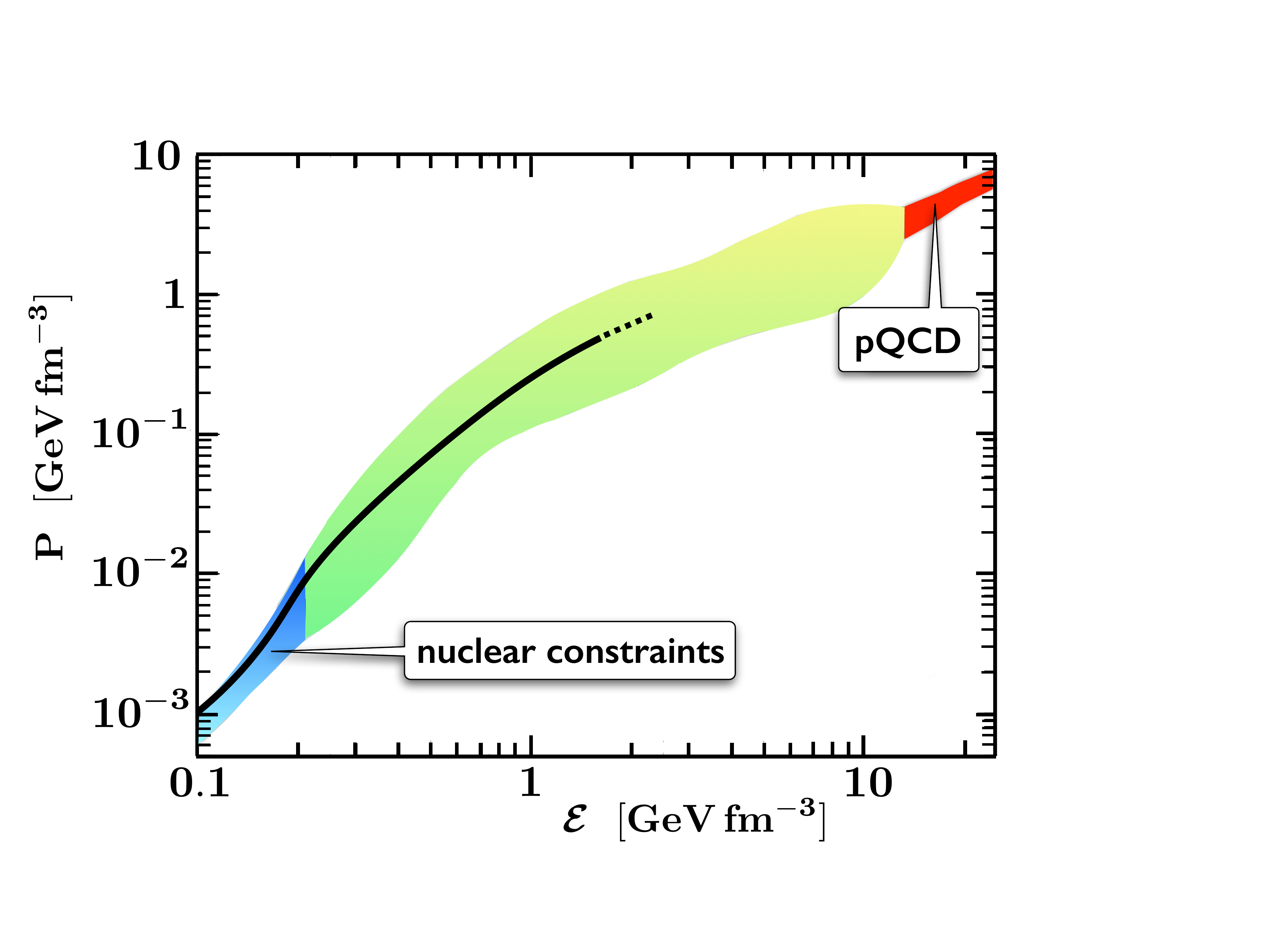}
\caption{Pressure as a function of energy density for neutron star matter with constraints from perturbative QCD and nuclear EFT calculations at the upper and lower ends of the density scale, adapted from Refs.\cite{Annala2018,Vuorinen2018}. The area between the upper and lower limits marks the range of acceptable equations of state subject to empirical conditions on neutron star maximum mass and tidal deformability, the latter from LIGO and Virgo gravitational wave analysis. The solid curve represents the chiral FRG equation of state $P({\cal E})$ as in Fig.\,\ref{fig:1}. }
\label{fig:3}
\end{center}
\end{figure*}

\subsection{Connection to relativistic Fermi-liquid theory and quasiparticles}

The observation that chiral nucleon-meson field theory in combination with FRG appears to successfully describe neutron star matter at core densities suggests a study of the Fermi-liquid properties of neutron matter in a relativistic framework \cite{BC1976}.

Given the neutron star EoS of the chiral FRG model, the aim is to deduce the spin-independent Fermi-liquid parameters $F_0$ and $F_1$ and examine their density dependence. The speed of sound (\ref{eq:sound-speed-Landau}) involves a combination of these Landau parameters. In order to determine
$F_0(\rho)$ and $F_1(\rho)$ separately, additional information on the quasiparticle properties is required. In the core of neutron stars the relevant quasiparticles are dominantly neutrons dressed by the strong interactions with the surrounding matter, plus a small fraction of protons. Once again, we assume that the small proton fraction in neutron star matter can be neglected in the analysis of the spin-independent Fermi liquid properties.

The single-particle motion in a relativistic theory of neutron matter is described by the in-medium Dirac equation \cite{Brockmann:1996xy}
\begin{equation}
\left[\slashed{p}-M_0-\Sigma_s(p;\rho)-\slashed{\Sigma}_v(p;\rho)\right]u(\mathbf{p},s)=0,
\end{equation} 
where in standard notation $\slashed{p}=\gamma_\mu \,p^\mu$, $p^\mu=(p_0,\mathbf{p})$, $\Sigma_s$ is the scalar, $\Sigma_v^\mu=(\Sigma_0,\mathbf{\Sigma})$ is the vector neutron self energy, and $u(\mathbf{p},s)$ is the corresponding Dirac spinor. The quasiparticle energies (in the rest frame of the system) are defined by the self-consistent solutions of the equation \cite{vanDalen:2005ns}
\begin{equation}\label{eq:qp-energy-full}
\varepsilon_p=\sqrt{\mathbf{p}^2+[M_0+\Sigma_s(\mathbf{p},\varepsilon_p;\rho)]^2}+\Sigma_0(\mathbf{p},\varepsilon_p;\rho) .
\end{equation}
Here, the spatial component of the vector self-energy $\mathbf{\Sigma}$ vanishes as in the mean-field treatment discussed above.
In order to obtain a tractable scheme, we set $|\mathbf{p}|=p_F$ in the self-energies and use the following {\em ansatz}
for the quasiparticle energy
\begin{equation}\label{eq:qp-energy-ansatz}
\varepsilon_p=\sqrt{\mathbf{p}^2+M(\rho)^2}+U(\rho) ,
\end{equation}
with a density-dependent Fermion (nucleon) mass $M(\rho)=M_0+\Sigma_s(p_F,\varepsilon_F;\rho)$ and an effective vector potential $U(\rho)=\Sigma_0(p_F,\varepsilon_F;\rho)$. 
In Sec.\, \ref{sec:results} we discuss possible implications of this approximation. 

Consider again the Fermi velocity \cite{BC1976},
\begin{equation}
v_F = \left(\frac{\partial\varepsilon_p}{\partial p}\right)_{p=p_F} = 
\frac{p_F }{ \mu\left(1+\frac13 F_1\right)}~.
\label{eq:vf1}
\end{equation} 
Thus the Fermi velocity as a function of baryon density $\rho$ yields $F_1(\rho)$. The parameter $F_0(\rho)$ is then obtained by insertion into Eq.\,(\ref{eq:sound-speed-Landau}) (cf.~Eq.~(10) in \cite{Landau2}):
\begin{equation}
F_0(\rho) = \frac{3\mu}{ p_F\,v_F}\,c_1^2(\rho) -1~.
\label{eq:F0}
\end{equation}
A key quantity for further steps is obviously the baryon chemical potential $\mu(\rho)$ as a function of baryon density. This is obtained by first constructing the energy density ${\cal E}(\rho)$ as a function of $\rho$. The (zero temperature) thermodynamic relation
\begin{equation}
{\cal E} + P({\cal E}) = \rho\frac{\partial{\cal E}}{\partial\rho}~,
\end{equation}
yields the density as a function of ${\cal E}$
\begin{equation}
\rho({\cal E}) = \rho^{(0)}\exp\left[\int_{{\cal E}^{(0)}}^{\cal E}\frac{d{\cal E}'}{{\cal E}' +P({\cal E}')}\right]~.
\label{eq:xdensity}
\end{equation}
Here the lower limit of the integral is chosen at a very low density $\rho^{(0)}$, where the energy density is well approximated by the mass density,
\begin{equation}
{\cal E}^{(0)} \equiv {\cal E}(\rho^{(0)})= M_0\,\rho^{(0)}~.\nonumber
\end{equation}
By inverting Eq.\,(\ref{eq:xdensity}) with $P({\cal E})$ as input, we obtain the energy density ${\cal E}(\rho)$ and also the baryon chemical potential, $\mu = \partial{\cal E}/\partial\rho$, as functions of the density. 
With our {\em ansatz} (\ref{eq:qp-energy-ansatz}) for the quasiparticle energy, the Fermi energy is given by
\begin{equation}\label{eq:qp-ansatz}
\mu = \varepsilon_{p=p_F} = \sqrt{p_F^2 + {M}{^2}(\rho)} + U(\rho)~.
\end{equation}
It follows that
\begin{equation}
v_F(\rho) = \frac{p_F}{\sqrt{p_F^2 + {M}{^2}(\rho)}}~.
\label{eq:vf2}
\end{equation}

In terms of the energy per particle, $E(\rho)/A$, the energy density is given by:
\begin{equation}
{\cal E}(\rho) = \left(M_0 +\frac{E(\rho)}{ A}\right)\rho~,
\label{eq:energyden}
\end{equation}
and the chemical potential by
\begin{equation}
\mu = M_0 + \left(1 + \rho\frac{\partial}{\partial\rho}\right)\frac{E(\rho)}{ A}~.
\label{eq:chempot}
\end{equation}
Equating the two expressions for $\mu$, we find
\begin{equation}
M_0 + \left(1 + \rho\frac{\partial}{\partial\rho}\right)\frac{E(\rho)}{ A} =  m^*(\rho) + U(\rho)~.
\label{eq:chempot-eq}
\end{equation}
where we have identified the Landau effective mass 
\begin{equation}\label{eq:Landau-mstar}
m^*(\rho) = \sqrt{p_F^2 + {M}{^2}(\rho)} .
\end{equation}
Thus, the density dependence of $E(\rho)/A$ is reflected in the density-dependent mass $M(\rho)$ and the potential $U(\rho)$. 

In the chiral FRG model, the in-medium nucleon mass $M(\rho)$ scales with the expectation value of the scalar $\sigma$ field, which is identified with the in-medium pion decay constant $f_\pi^*(\rho)$, associated with the time component of the axial current. This plays the role of a chiral order parameter. Consequently, in this model we have: 
\begin{equation}
M(\rho) = M_0\,\frac{\langle\sigma\rangle}{ f_\pi}= M_0\,\frac{f_\pi^*(\rho)}{ f_\pi}~.
\end{equation}
The proportionality of the nucleon mass to the pion decay constant originates from current algebra and the Goldberger-Treiman relation. The chiral FRG approach maintains this property for a nucleon in a dense medium.  

Finally, given $M(\rho)$ and $U(\rho)$ one can return to Eqs.\,(\ref{eq:xF0},\ref{eq:xF1}) or (\ref{eq:yF0},\ref{eq:yF1}) and determine the Landau parameters $F_0$ and $F_1$.

\section{Results: Quasiparticles and Fermi-Liquid parameters}\label{sec:results}

\subsection{Bulk and quasiparticle properties}

Using Eq.\,(\ref{eq:xdensity}), with the chiral FRG equation of state $P({\cal E})$ as input, we compute the energy density, ${\cal E}(\rho)$, and the energy per particle, shown in Figs.\,\ref{fig:4} and \ref{fig:5}. The corresponding baryon chemical potential $\mu(\rho)= \partial{\cal E}/\partial\rho$ is shown in Fig.\,\ref{fig:6}. 

\begin{figure*}[t]
\begin{center}
\includegraphics[height=60mm,angle=-00]{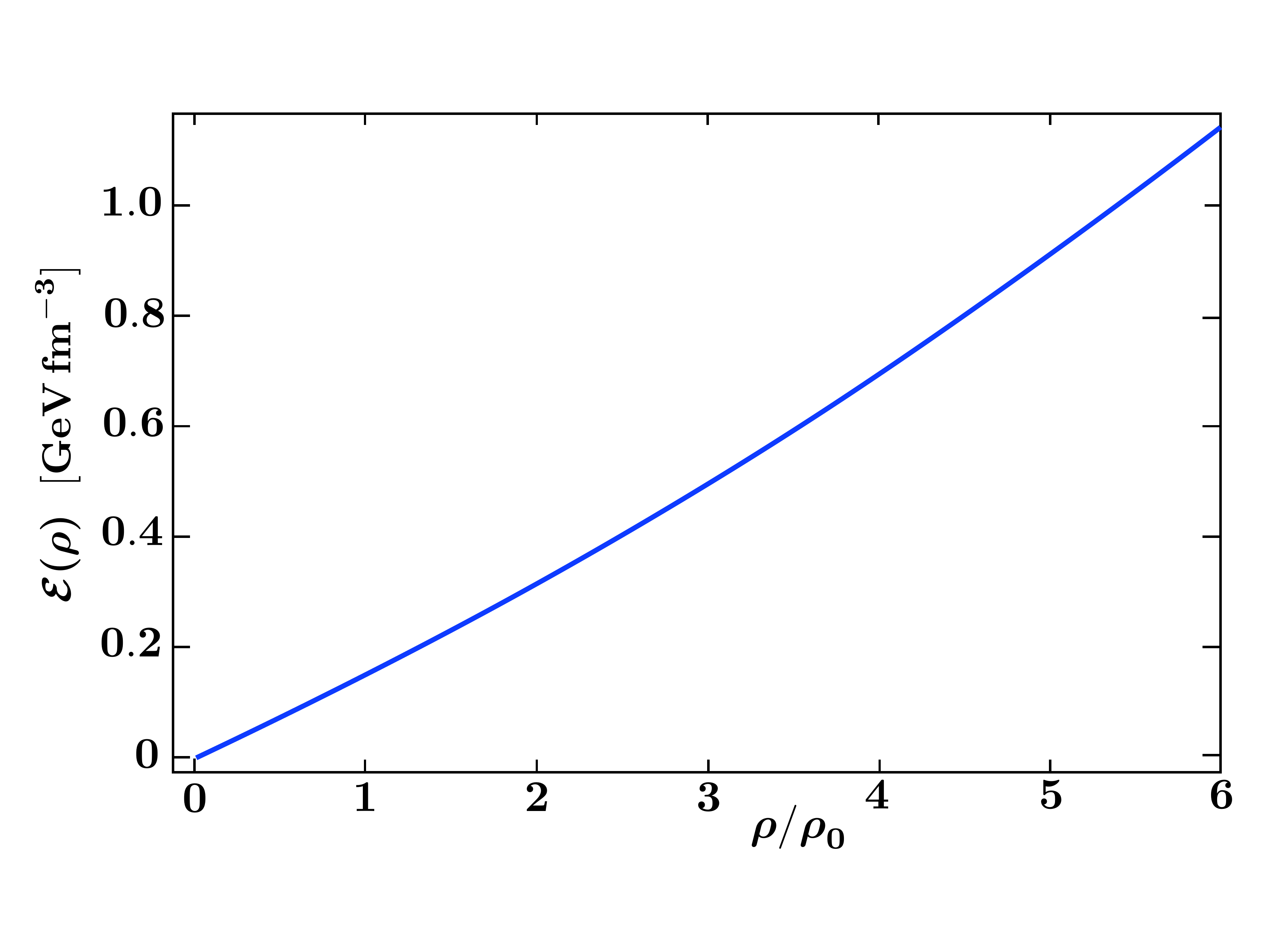}
\caption{Energy density of neutron star matter from chiral FRG calculations \cite{DW2015,DW2017} as a function of density given in units of $\rho_0 = 0.16$ fm$^{-3}$.}
\label{fig:4}
\end{center}
\end{figure*}

\begin{figure*}[t]
\begin{center}
\includegraphics[height=60mm,angle=-00]{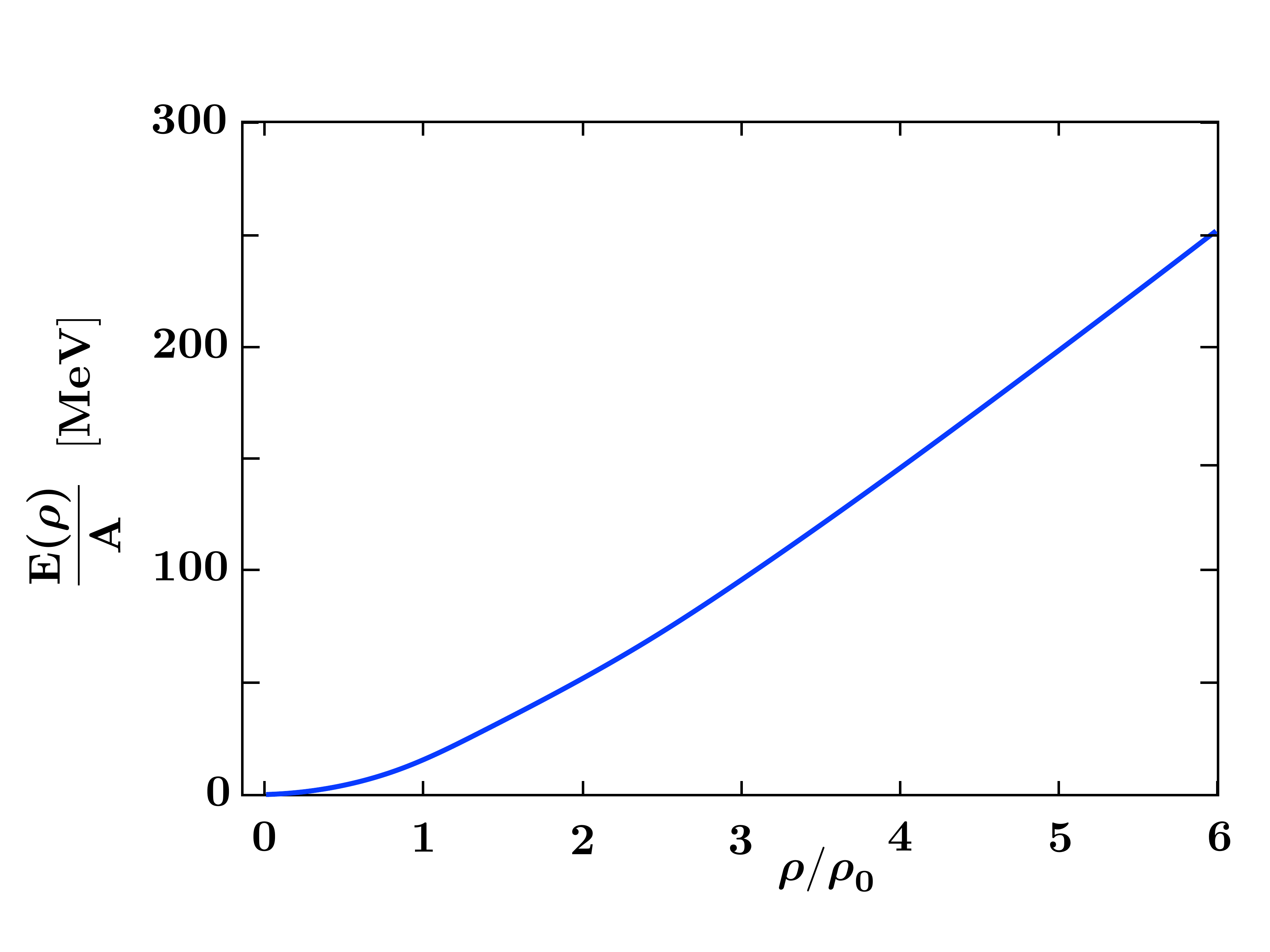}
\caption{Energy per particle deduced from the chiral FRG equation of state as a function of baryon density given in units of $\rho_0 = 0.16$ fm$^{-3}$.}
\label{fig:5}
\end{center}
\end{figure*}

\begin{figure*}[t]
\begin{center}
\includegraphics[height=65mm,angle=-00]{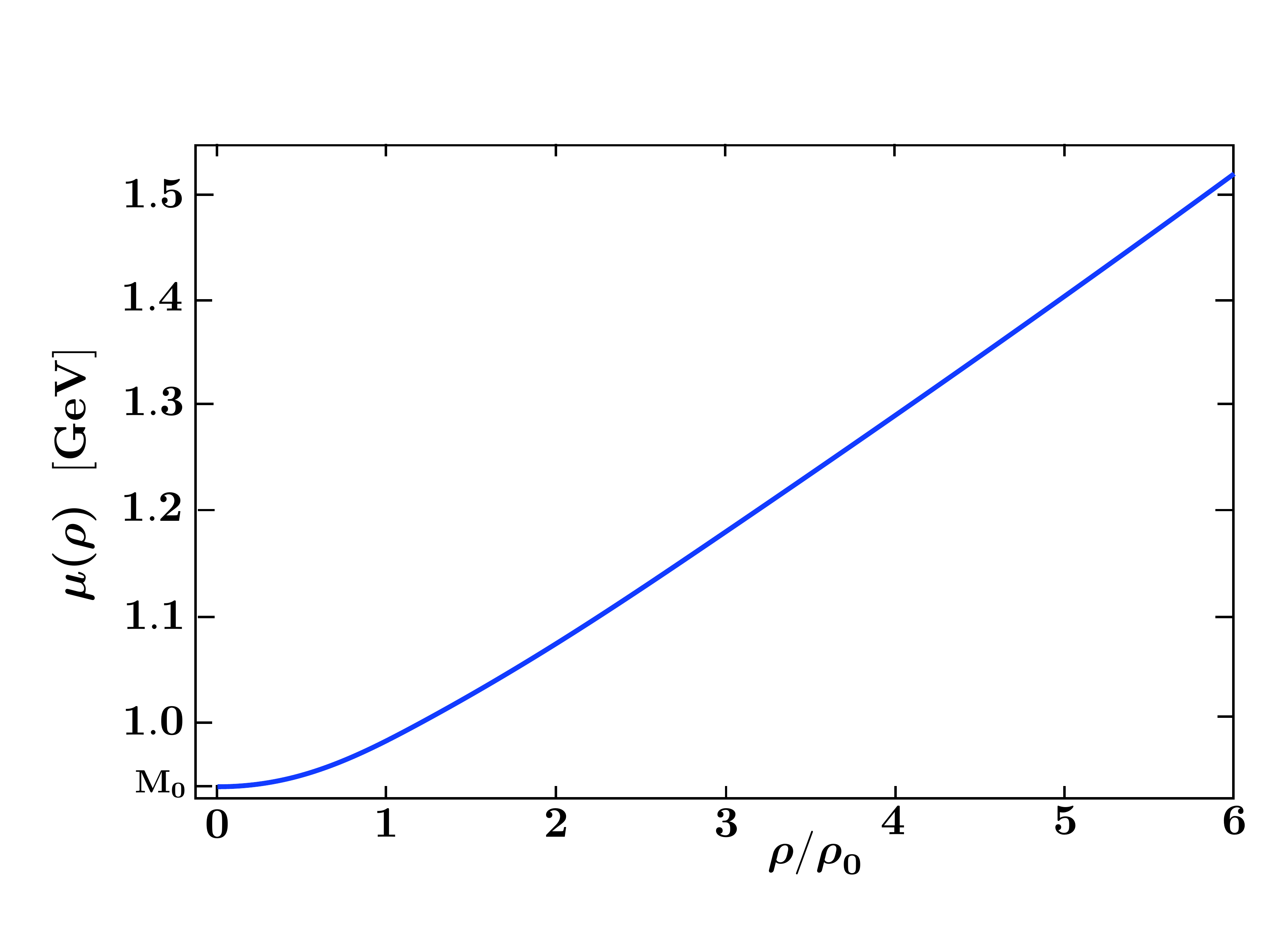}
\caption{Chemical potential of neutron matter from ChNM-FRG calculations \cite{DW2015,DW2017} as a function of density given in units of $\rho_0 = 0.16$ fm$^{-3}$. }
\label{fig:6}
\end{center}
\end{figure*}

The next step is the decomposition of the baryon chemical potential (\ref{eq:chempot}) in terms of the density-dependent quasiparticle mass and effective potential. As mentioned, the mass $M(\rho)$ scales with the density dependence of the chiral order parameter $\langle\sigma\rangle = f_\pi^*(\rho)$. Even at $\rho\approx 5\,\rho_0$, i.e., at densities encountered in the inner core region of neutron stars, the in-medium nucleon mass is still almost half of its vacuum value. This indicates that the spontaneous breaking of the chiral symmetry remains strong at such large densities. Here the nonperturbative treatment of fluctuations beyond the mean-field approximation in the chiral FRG approach is of crucial importance \cite{DW2015, DW2017}. In a mean-field calculation one finds instead a first-order chiral phase transition at densities as low as $\rho\approx 3\,\rho_0$. With inclusion of fluctuating fields and, in particular, many-body correlations featuring repulsive Pauli effects that become increasingly active with increasing density, the transition to chiral symmetry restoration is shifted to densities beyond $5\,\rho_0$.

Once $M(\rho)$ is determined (see Appendix C for numerical details), the effective potential is obtained by
\begin{equation}
U(\rho) = \mu - m^*(\rho)~,
\end{equation}
with the Landau effective mass $m^*(\rho)$, Eq.~(\ref{eq:Landau-mstar}).
Results for the in-medium mass $M(\rho)$ and the effective potential $U(\rho)$ are summarized in Fig.\,\ref{fig:7}. The deviations in $M(\rho)$ from linear density dependence reflect correlations and fluctuations beyond the mean field approximation. The Landau effective mass, $m^*(\rho)$, is shown in Fig.\,\ref{fig:8}. This is the quantity that determines the Landau parameter $F_1$, see Eq.\,(\ref{eq:yF1}). At large densities, the neutrons at the Fermi surface become relativistic, with an effective mass $M(\rho)$ comparable to or smaller than $p_F$. Nevertheless, at the densities relevant to neutron stars, $M(\rho)$ is still large compared to values expected when approaching chiral symmetry restoration. 

It is important to point out that $M(\rho)$ and $U(\rho)$ behave quite differently in comparison with the corresponding quantities of standard relativistic mean-field (RMF) models, which yield much stronger scalar and vector mean fields. This difference comes primarily from the explicit treatment of multipion exchange processes in the chiral FRG theory. For example, we have $U(\rho = \rho_0) \approx 0.14$ GeV whereas the corresponding RMF vector potential would be almost three times as large. Similarly, the density-dependent mass $M(\rho)$ in the chiral FRG approach drops to $M(\rho=\rho_0) \approx 0.83\,M_0$ at normal nuclear matter density. The equivalent attractive scalar potential is $U_s(\rho=\rho_0)\equiv M(\rho_0)-M_0 \approx -0.16$ GeV, whereas the typical scalar potential in RMF models is about twice as strong. In the chiral FRG scheme, much of the intermediate-range attraction between nucleons in the medium is generated by two-pion exchange processes treated explicitly, with inclusion of Pauli effects. 

\begin{figure*}[t]
\begin{center}
\includegraphics[height=70mm,angle=-00]{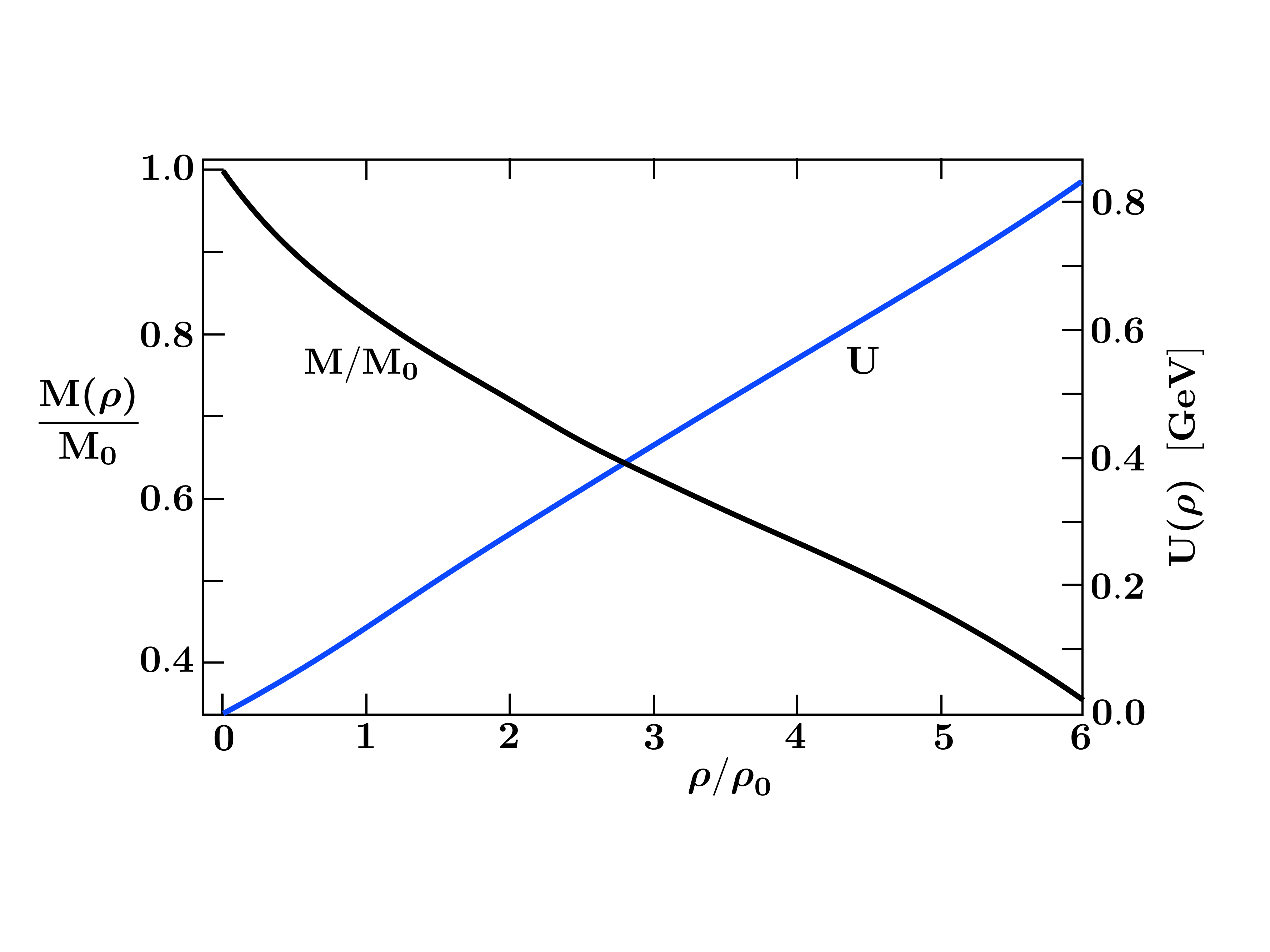}
\caption{Neutron quasiparticle mass $M(\rho)$ and effective potential $U(\rho)$ as functions of density. The effective mass is given in units of the vacuum neutron mass, $M_0 = 939.57$ MeV. }
\label{fig:7}
\end{center}
\end{figure*}

\begin{figure*}[t]
\begin{center}
\includegraphics[height=60mm,angle=-00]{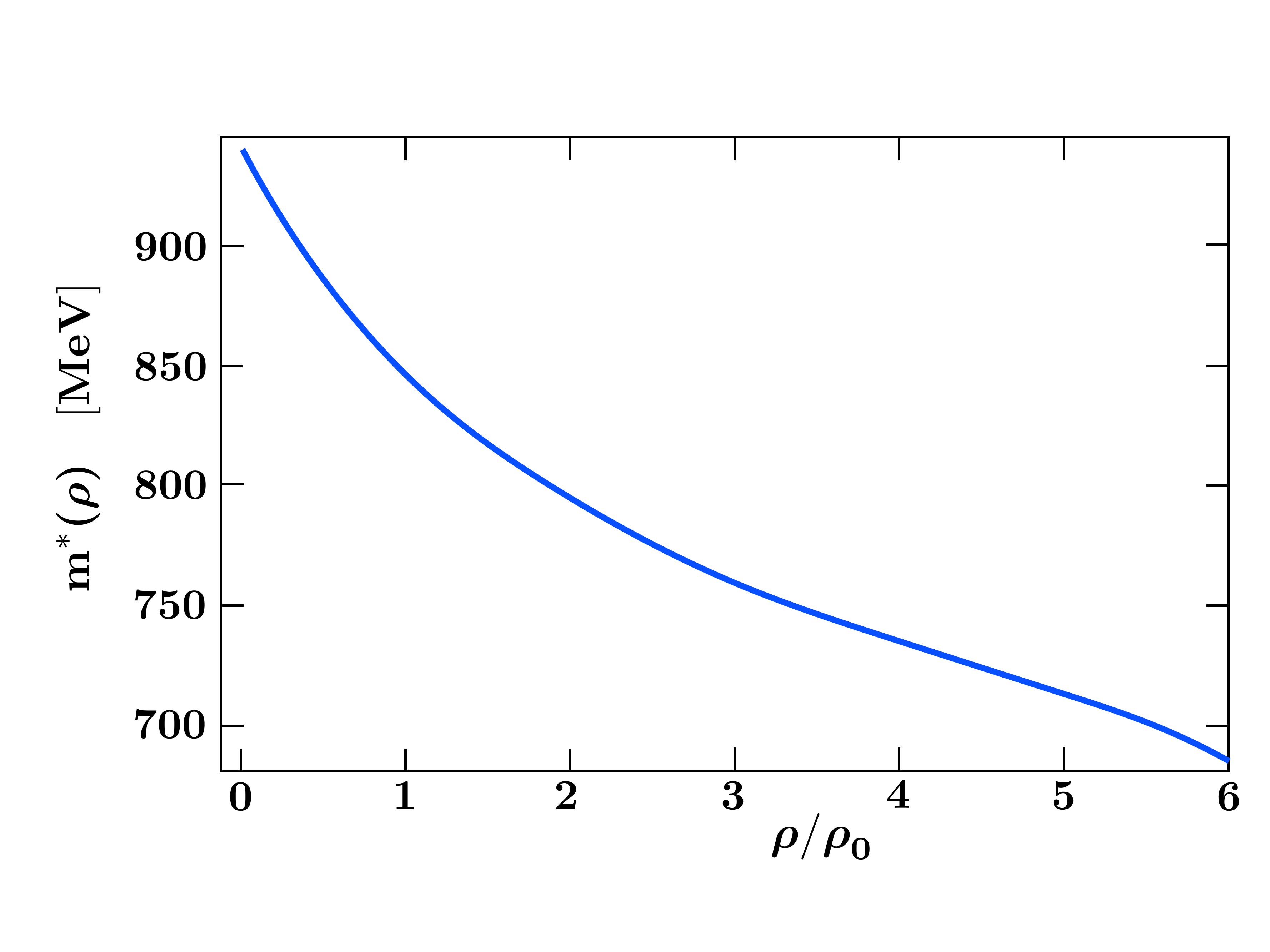}
\caption{Landau effective mass $m^*(\rho) = \sqrt{p_F^2 + M^2(\rho)}$ as a function of density in units of $\rho_0 = 0.16$ fm$^{-3}$.}
\label{fig:8}
\end{center}
\end{figure*}

\begin{figure*}[t]
\begin{center}
\includegraphics[height=70mm,angle=-00]{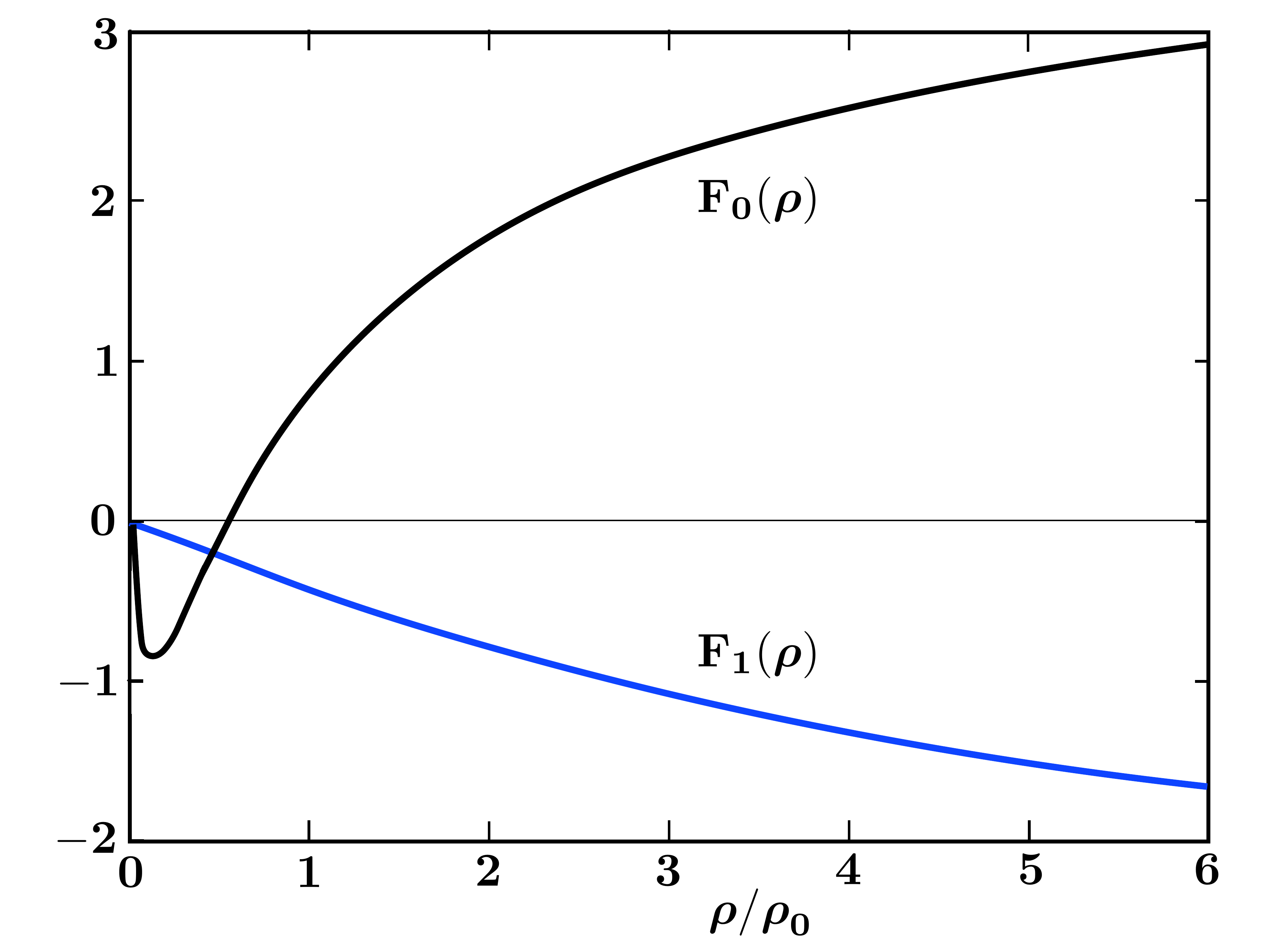}
\caption{Landau Fermi liquid parameters $F_0$ and $F_1$ for neutron star matter as a function of density $\rho/\rho_0$ in units of $\rho_0 = 0.16$ fm$^{-3}$.}
\label{fig:9}
\end{center}
\end{figure*}

\subsection{Landau parameters \texorpdfstring{$F_0$}{F0} and \texorpdfstring{$F_1$}{F1}}

The input is now prepared to evaluate the dimensionless Fermi liquid parameters $F_0$ and $F_1$ as functions of density, using Eqs.\,(\ref{eq:xF0} and \ref{eq:xF1}). The results for these Landau parameters are plotted in Fig.\,\ref{fig:9}. 

At low densities, $\rho < \rho_0$, the parameter $F_0$ starts out negative and then turns positive around $\rho_0$. This behavior is consistent with a perturbartive ChEFT calculation for neutron matter \cite{HKW2013}. With next-to-next-to-next-to-leading order ($N^3LO$) chiral NN interactions and $N^2LO$ three-body forces, a second-order many-body calculation gives $f_0(\rho=0.2\,\rho_0) = - 1.25$ fm$^{-2}$ and $f_0(\rho=\rho_0) = 0.7$ fm$^{-2}$. The resulting dimensionless $F_0 = N(0)f_0$ with the density of states $N(0)= m^*p_F/\pi^2$ is indeed close to the one deduced from the chiral FRG EoS at low densities. 

On the other hand the Landau effective mass $m^*$ from the ChEFT calculation \cite{HKW2013} does not decrease as fast as the one found in the relativistic chiral FRG model. The resulting $F_1(\rho = \rho_0)$ is close to zero but with a negative slope indicating a decreasing effective mass at higher density, as observed in the chiral FRG result.  

Historically, the behavior of the nucleon effective mass close to the Fermi surface has in fact been subject to many detailed investigations. A representative overview can be found in Ref.\,\cite{Mahaux1985}. While $m^*$ in mean-field calculations is generally less than the free mass, correlations involving particle-hole excitations tend to increase the effective mass near the Fermi surface. Thus, at least at densities $\rho\lesssim \rho_0$, the resulting Landau effective mass in neutron matter remains close to its vacuum value \cite{HKW2013,Ismail:2019rjg}. Our parametrization of the quasiparticle energy (\ref{eq:qp-energy-ansatz}) is presumably not able to fully capture these detailed correlation effects. Consequently, at low densities we find that the Landau effective mass, and consequently also $F_1$, is somewhat smaller than obtained in ChEFT \cite{HKW2013}. 

A more detailed treatment of the correlations near the Fermi surface, e.g., retaining the explicit momentum dependence of the neutron self-energies in (\ref{eq:qp-energy-full}), is expected to yield an $F_1$ somewhat larger than in the present analysis \cite{vanDalen:2005ns,HKW2013}. Consequently the $F_0(\rho)$ shown in Fig.\,\ref{fig:9} is presumably a lower bound. 

The overall strong increase of $F_0$ at high densities reflects the growing importance of repulsive many-body correlations as the matter gets more and more compact. Part of this effect is due to the action of the Pauli principle on nucleons fluctuating around the Fermi surface. Such repulsive correlations are at the same time responsible for the increase of the sound velocity at high densities beyond its canonical value, $c_1 > 1/\sqrt{3}$. 

\subsection{Upper bound for \texorpdfstring{$F_0$}{F0}}

So far we have assumed that the density-dependent neutron mass, $M(\rho)$, is proportional to the in-medium pion decay constant, $f_\pi^*(\rho)$, and decreases continuously with increasing density. In order to assess uncertainties possibly implied by this assumption, it is instructive to examine a limiting case, replacing $M(\rho)$ by the constant free neutron mass $M_0$ at all densities. Given the speed of sound (\ref{eq:sound-speed-Landau}) constrained by astrophysical observations, such an extreme limit reduces the magnitude of (negative) $F_1 <0$, balanced by a corresponding increase of $F_0$. The resulting $F_0$ can be considered as an upper limit that provides an estimate of uncertainties in the determination of the Landau parameters.

With the constant (vacuum) mass $M_0 = g f_\pi$ as input, the chemical potential is 
\begin{equation}\label{eq:chem-pot-free-mass}
\mu = \sqrt{p_F^2 + M_0^2} + V_0(\rho)\,,
\end{equation}
where the previously used vector potential $U(\rho)$ in Eq.\,(\ref{eq:qp-ansatz}) is now replaced by $V_0(\rho)$ subject to the condition that $\mu = \partial{\cal E}/\partial\rho$ remains unchanged, given by the values listed in Table \ref{t1} and shown in Fig.\,\ref{fig:6}. This limiting vector potential is plotted in Fig.\,\ref{fig:10}. Note that $V_0(\rho)$ is weaker in magnitude than $U(\rho)$ because the attraction, previously manifest in the decreasing $M(\rho)$, is now effectively transferred to $V_0$. 
Also shown in Fig.\,\ref{fig:10} is the Landau effective mass divided by the chemical potential,  $m^*/\mu = \sqrt{p_F^2 + M_0^2}/\mu = 1-V_0/\mu$. This ratio is close to unity at densities up to about $2\rho_0$. Hence $F_1$ stays close to zero within that density range, in qualitative agreement with the results obtained in ChEFT~\cite{HKW2013}.

\begin{figure*}[t]
\begin{center}
\includegraphics[height=70mm,angle=-00]{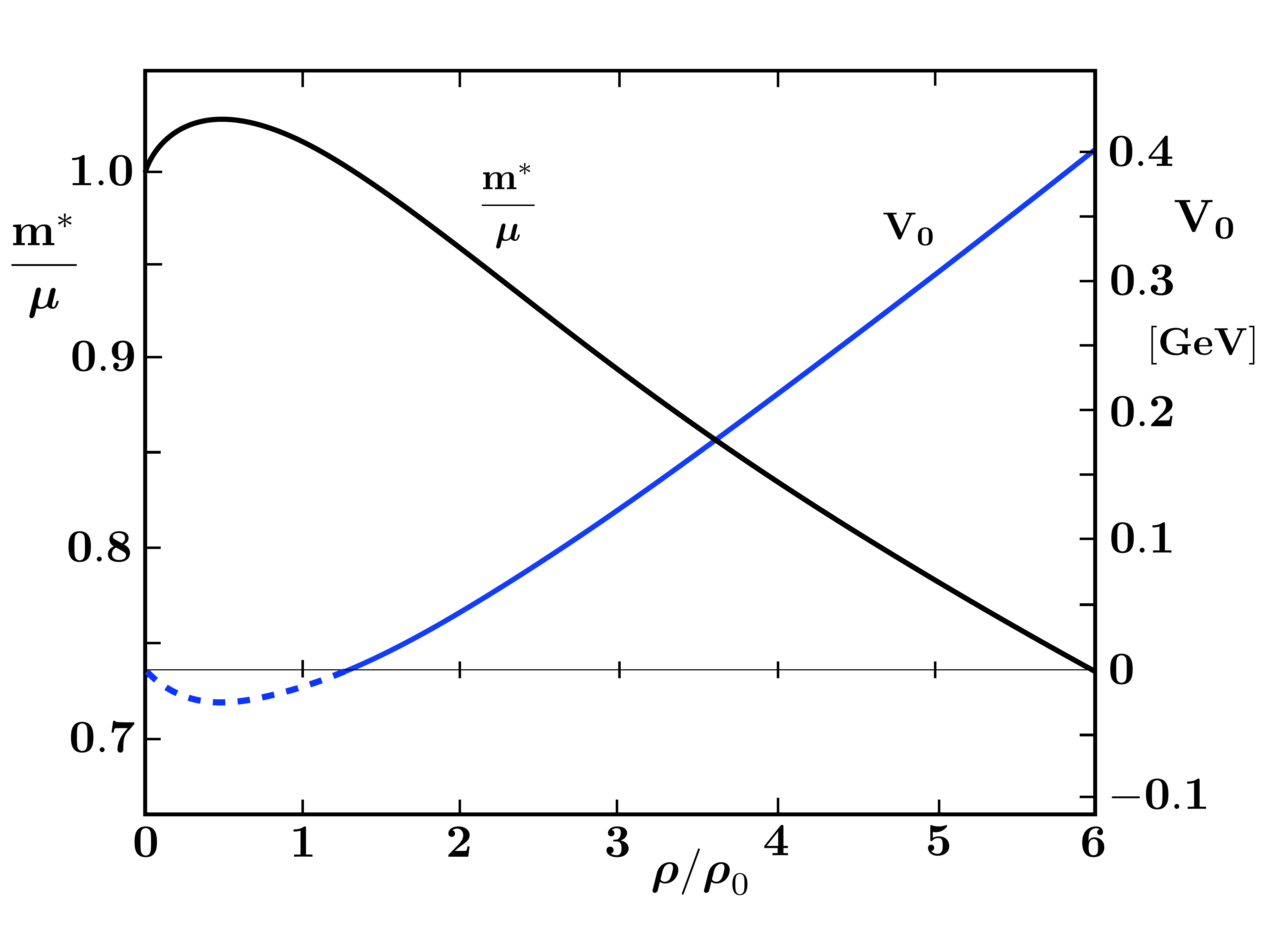}
\caption{The effective vector potential $V_0(\rho)$ and the ratio of Landau effective mass over chemical potential, $m^*/\mu$, for the limiting case $M(\rho)\equiv M_0$, as functions of baryon density in units of $\rho_0 = 0.16$ fm$^{-3}$. The dashed line indicates the range of densities, where the {\em ansatz} (\ref{eq:chem-pot-free-mass}) yields $V_0<0$, at variance with the model assumptions (see main text). }
\label{fig:10}
\end{center}
\end{figure*}

\begin{figure*}[t]
\begin{center}
\includegraphics[height=70mm,angle=-00]{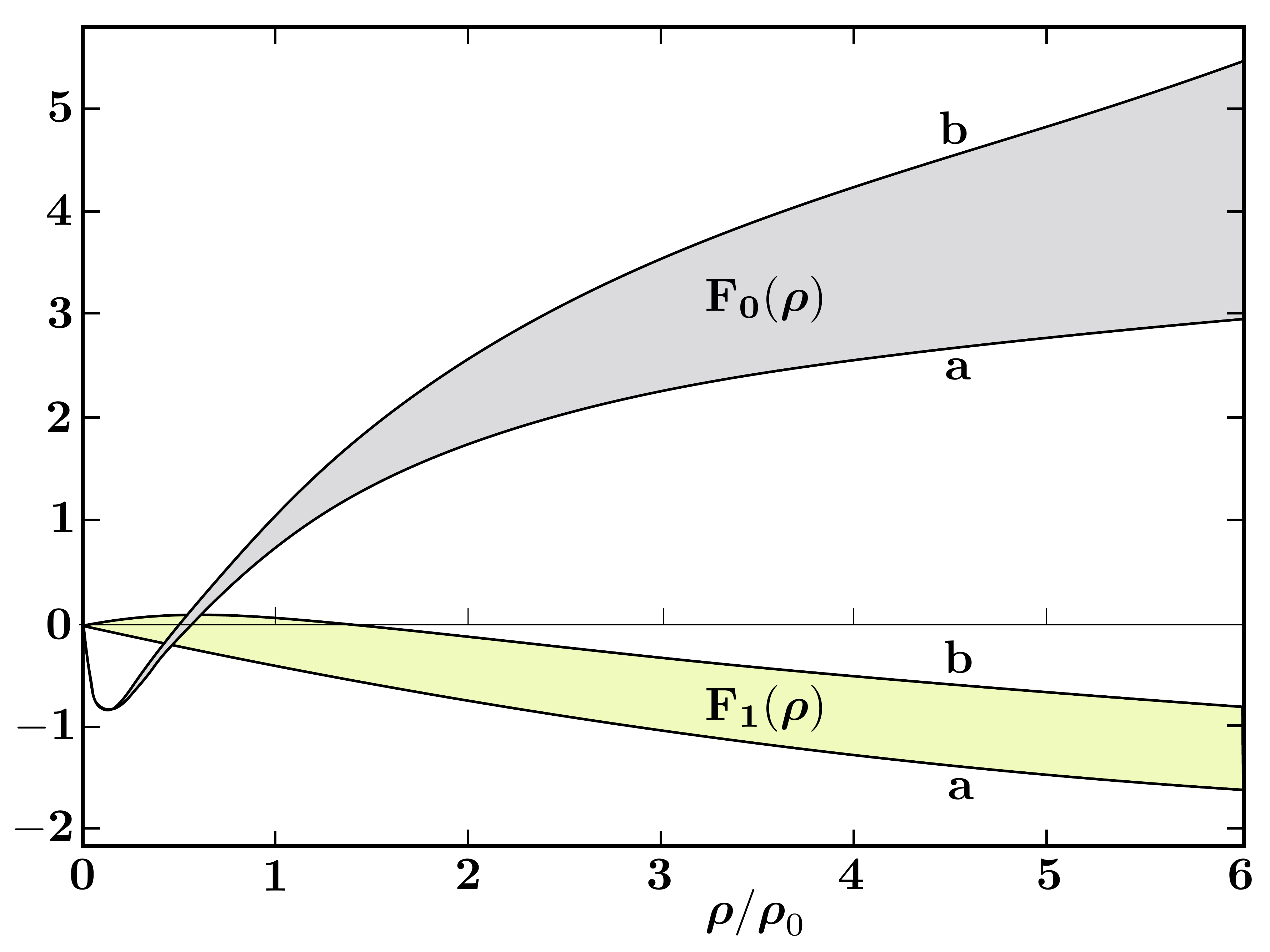}
\caption{Landau parameters $F_0$ and $F_1$ as functions of baryon density in units of $\rho_0 = 0.16$ fm$^{-3}$. The upper and lower bounds of the shaded areas correspond to the following limiting cases for choices of the baryon mass: (a) lower boundary lines: $M(\rho)/M_0 = \langle\sigma\rangle(\rho)/f_\pi$ as in Fig.\,\ref{fig:9}; (b) upper boundary lines: $M(\rho) \equiv M_0$ at all densities. }
\label{fig:11}
\end{center}
\end{figure*}

The Landau parameters  are given by [see Eqs.\,(\ref{eq:xF0}) and (\ref{eq:xF1})]:
\begin{equation}
F_0(\rho) = {p_F\over\pi^2} \sqrt{p_F^2 + M_0^2}~ {\partial V_0(\rho)\over\partial\rho}~~,~~~~F_1(\rho) = -{3 V_0(\rho)\over \mu(\rho)}~.
\end{equation}
They are shown in Fig.\,\ref{fig:11} together with the ones previously computed. The areas between the upper and lower boundary curves for $F_0$ and $F_1$ can be considered as uncertainty measures covering the two limiting cases, $M(\rho) \propto f_\pi^*(\rho)$ (lower bounds) and $M(\rho)\equiv M_0$ (upper bounds).

We note that at low densities, where the effective vector potential is negative, the {\em ansatz} (\ref{eq:chem-pot-free-mass}) is strictly speaking not consistent with the model assumptions, since the interaction between two baryons mediated by the exchange of an isoscalar vector boson is repulsive.\footnote{This inconsistency would presumably not occur in a more refined approach, where the momentum dependence of the potential $V_0$ is accounted for (see, e.g.,~\cite{vanDalen:2005ns}).} Nevertheless, as already indicated, the resulting values for the Landau effective mass at low densities are consistent with ChEFT. We can therefore conclude that our estimate of the upper bound on the Landau parameters remains valid also at low densities.

\subsection{Zero sound}

Cold Fermi liquids can develop a sound-like collective mode, {\it zero sound} \,\cite{Landau2}. The velocity of zero sound,
$c_0 = \omega/q$ (in terms of frequency and wave vector of the mode), is yet another characteristic property of the fluid that is linked to $F_0$ and $F_1$. In particular, for the model considered, where $F_\ell=0$ for $\ell \geq 2$, the zero-sound velocity is determined by real solutions of the equation \cite{Abrikosov1959,BP1991}
\begin{equation}
\label{eq:zerosound}
\left(F_0 + {F_1\over 1+ F_1/3}\,s^2\right)\Omega_{00}(s) = -1~,
\end{equation}
where 
\begin{equation}
\Omega_{00}(s) = 1+{s\over 2}\ln\left({s-1\over s+1}\right)
\end{equation}
is the long-wavelength limit of the Lindhard function \cite{Lindhard:1954va} \footnote{Note that in Ref.\cite{Matsui1981} $\Phi(s) = - \Omega_{00}(s)$ is used.} and $s = c_0/v_F$, the velocity of the sound mode divided by the Fermi velocity. 

The zero sound velocity in units of the velocity of light is shown in Fig.\,\ref{fig:12}. A comparison is once again made between the standard case with density dependent mass $M(\rho) = g\langle\sigma\rangle$ and the limit $M(\rho) \equiv M_0$. With the former choice it turns out that Eq.\,(\ref{eq:zerosound}) permits real solutions of $c_0$ only in a restricted range of densities $\rho$. Outside this range the zero-sound velocity is complex. Its imaginary part indicates Landau damping.

From Fig.\,\ref{fig:12} one concludes that, while neutron star matter becomes a relativistic fluid at high densities, it strictly satisfies the causality constraint, $c_0 < c$. This is implied by the strong repulsion encoded in the vector field that grows linearly at high density and contributes to both $F_0$ and $F_1$. Notably, setting $F_1= 0$ would not be consistent as it would lead to a superluminal velocity of zero sound at high densities.

\begin{figure*}[t]
\begin{center}
\includegraphics[height=70mm,angle=-00]{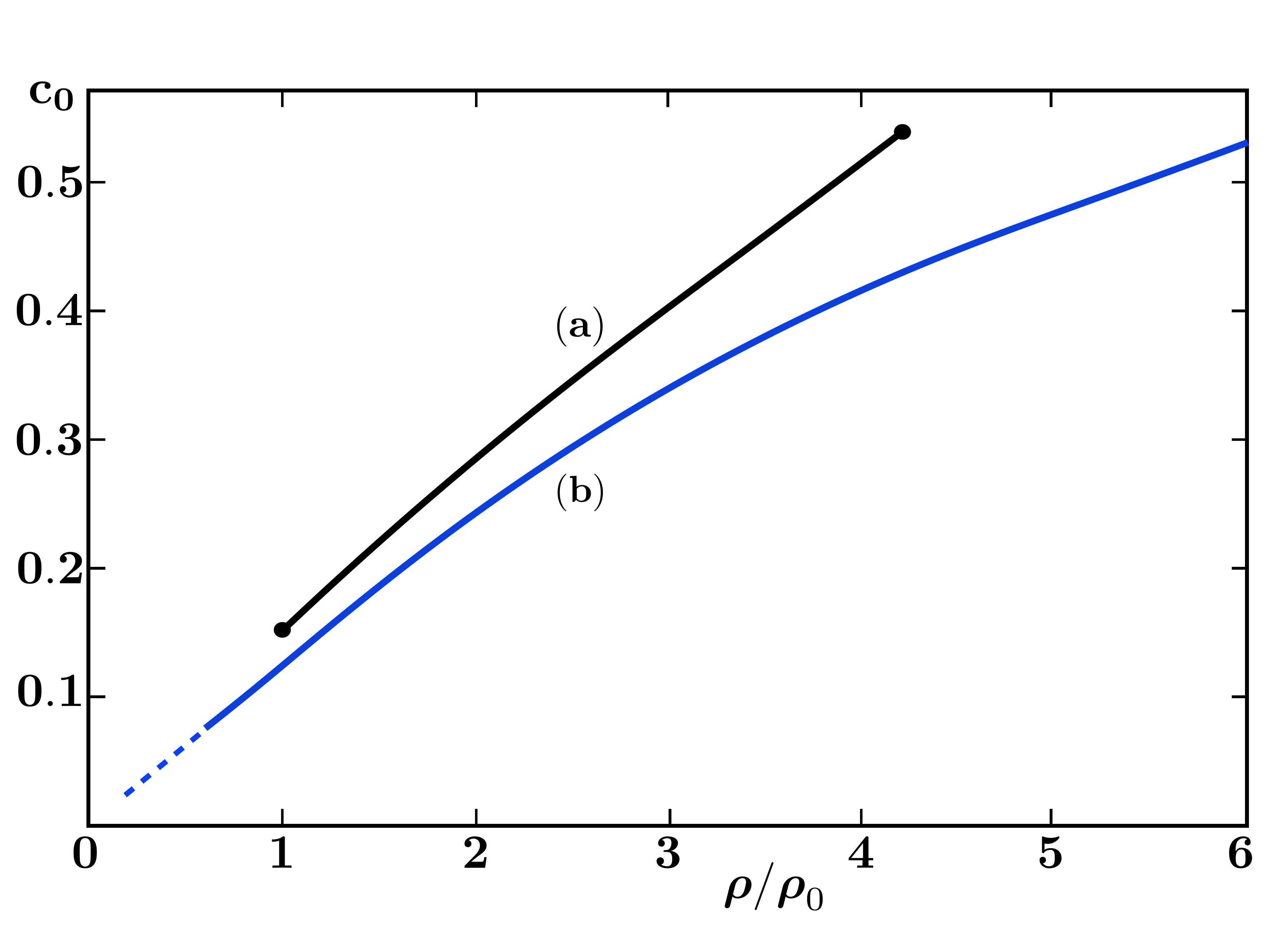}
\caption{Velocity $c_0$ (in units of the speed of light) of the zero-sound collective mode, as a function of baryon density in units of $\rho_0 = 0.16$ fm$^{-3}$.  The two curves correspond to different choices of quasiparticle masses. Curve (a): baryon mass $M(\rho) \propto f_\pi^*(\rho)$; curve (b): using $M(\rho)\equiv M_0$. At densities below and above the end points indicated in curve (a) the zero-sound velocity is complex.}
\label{fig:12}
\end{center}
\end{figure*}

\subsection{Comparison with liquid helium-3}

By their magnitudes, the Landau parameters are measures of the correlation strength in the Fermi liquid. At this point a comparison with another example of a strongly correlated fermionic many-body system, liquid $^3$He, is quite instructive \cite{Vollhardt1984,Leggett2016}. Normal $^3$He at low temperature is a high-density liquid in which the average distance between the helium atoms is of the same order as the atomic diameter. Their interaction has an attractive van der Waals part with a range of a few angstroms, and a strongly repulsive short-range core. Apart from a re-scaling of distances by a factor of $10^{-5}$ between {\AA} and fm, this is qualitatively reminiscent of the situation in neutron matter at densities $\rho \approx 5\,\rho_0$, where the average distance between baryons is of the same order as the diameter of their compact valence quark cores, and the interaction is also characterized by the combination of intermediate-range attraction and strong short-range repulsion. The dimensionless Landau parameters $F_0$ and $F_1$ can thus give, by  comparison, an impression of how strong the correlations are in these two systems. 

A qualitative difference between liquid $^3$He and neutron matter is seen in the quasiparticle effective masses.  The Landau effective mass $m^*$ in  liquid $^3$He is a factor of approximately 3 to 6 larger (depending on pressure) than the mass of an isolated $^3$He atom \cite{Wheatley1975,Greywall1986}, indicating the presence of strongly repulsive correlations. This implies that $F_1$ for liquid $^3$He is positive and increasing with pressure, in contrast to the much more moderate modifications of $m^*$ with increasing pressure in neutron matter. Another qualitative distinction between the matter in neutron star cores and liquid $^3$He is the highly nonrelativistic nature of the latter. Characteristic Fermi velocities $v_F$ for $^3$He reported in a wide range of pressures from zero up to about 35 bar \cite{Wheatley1975} are on the order of $10^{-7}$ in units of the speed of light and actually decrease with increasing pressure as a consequence of the increasing effective mass. On the other hand, the sound velocity $c_1$ in liquid $^3$He is typically an order of magnitude larger than $v_F$, and their ratio
\begin{equation}
\left({c_1\over v_F}\right)^2 = {1\over 3}(1+F_0)(1+F_1/3)~,\nonumber
\end{equation}
is then reflected in large values of the Landau parameters. The following values are reported for liquid $^3$He (cf. Appendix C of Ref.\,\cite{BP1991}): at zero pressure, $F_0 \approx 9.3$ and $F_1 \approx 5.4$; at a pressure of $P = 27$ bar, $F_0 \approx 68.2$ and $F_1 \approx 12.8$. Understanding such large Fermi liquid parameters requires resummations to all orders and inclusion of collective modes in the quasiparticle interaction  (the induced interaction \cite{Babu-Brown1978}). We note that the neutron matter parameters $F_0$ and $F_1$ shown in Figs.\,\ref{fig:9} and \ref{fig:11} turn out to be strikingly smaller in magnitude than those for liquid $^3$He.

The reason for this qualitative difference can be traced to the scales at work in the respective quasiparticle interactions. Both neutron star matter and liquid $^3$He are governed by repulsive interactions with characteristic ranges, $r_c$. This correlation scale is to be compared with the average distance between Fermions in the medium $d\propto p_F^{-1}$. Their ratio $r_c/d\propto r_c\,p_F$ is a parameter that measures the growing importance of the repulsive forces as the density increases \cite{BM1969}. For neutron matter, even at densities several times larger than $\rho_0$, this parameter is still much smaller than in liquid $^3$He over a broad range of pressures. Thus, although the repulsive correlations in dense neutron matter are sufficiently strong to support two-solar-mass neutron stars against gravitational collapse, they appear as relatively moderate in comparison with those in liquid $^3$He. In this perspective, the dense baryonic matter encountered in the core of neutron stars is perhaps not as extreme as sometimes imagined.

\section{Summary and Concluding Remarks}

This work has focused on the Fermi liquid properties of dense baryonic matter as it may be realized in the center of neutron stars. While the detailed composition of strongly interacting matter at densities encountered in neutron star cores continues to be open to discussions of different scenarios ranging from conventional hadronic matter to quark matter, our starting point in this work has been an equation of state based on a chiral nucleon-meson field theory in combination with a nonperturbative approach using functional renormalization group methods. 

The input of the effective Lagrangian is tuned to pion-nucleon and pion-pion interactions in vacuum and to nuclear physics observables at densities around the equilibrium density of normal nuclear matter, $\rho_0\approx 0.16$\,fm$^{-3}$. The output is consistent with state-of-the-art chiral effective field theory calculations at baryon densities $\rho\lesssim 2\rho_0$. Moreover, the resulting EoS of neutron star matter is in agreement with observations, including the existence of two-solar mass stars and information from the gravitational wave signals produced by the merger of two neutron stars. 

It turns out that dense matter at zero temperature described by this EoS remains in the hadronic phase characterized by spontaneously broken chiral symmetry up to baryon densities even beyond five times the density of normal nuclear matter. This shifting of the chiral transition to extremely high densities, even above those encountered in neutron star cores, is a consequence of the important role played by fluctuations and correlations beyond mean-field approximation that are treated nonperturbatively in the FRG framework.

A special feature of this EoS is its stiffness, produced by repulsive multi-nucleon correlations that become increasingly important with increasing baryon density and drive the sound velocity significantly beyond the massless Fermi gas limit, $c_1^2 = 1/3$. Repulsive short-range correlations and the action of the Pauli principle in the dense medium are important ingredients behind this mechanism. It is then interesting to explore the quasiparticle properties of the dense Fermi system under such conditions.

Relativistic Fermi-liquid theory is applied to deduce the Landau parameters $F_0$ and $F_1$ and to investigate their density dependence. A relativistic treatment is mandatory because, at high densities, the Fermi momentum becomes large and comparable to the quasiparticle mass. The density-dependent neutron mass decreases in the compressed medium but still remains at about half of its vacuum value at densities typically reached in the neutron star core region. The magnitudes of the dimensionless Landau Fermi-liquid parameters then provide a measure of the many-body correlations as they grow continuously in strength with rising density.  Remaining uncertainties in $F_0$ are correlated with uncertainties in $F_1$, i.e., the quasiparticle effective mass and its density dependence. Thus, an estimate of the upper limit for $F_0$ is obtained by setting the in-medium nucleon mass equal to the free mass at all densities. Such a choice reduces the magnitude of $F_1$ and consequently increases $F_0$ under the condition of leaving the sound velocity unchanged. At densities up to $2\,\rho_0$, the resulting $F_1$ remains close to zero, in agreement with ChEFT results~\cite{HKW2013}. On the other hand, at higher baryon densities one may expect a more strongly negative $F_1$ if the chiral order parameter (the in-medium pion decay constant) is reduced from its vacuum value.

The results for the dimensionless parameters $F_0(\rho)$ and $F_1(\rho)$ display the typical behavior of a strongly correlated Fermi system. However, it is interesting to observe by comparison with Landau parameters in a system such as liquid $^3$He, that the correlations in neutron star matter are still fairly moderate. For example, while one finds $F_0 \approx 3$ at baryon densities $\rho\approx 0.8$ fm$^{-3}$ in neutron matter, the $F_0$ in normal liquid $^3$He is already more than three times larger even at zero pressure, and it strongly increases further with increasing pressure.   
Even in the extreme case of setting the in-medium baryon mass equal to its mass in vacuum, this Landau parameter does not exceed $F_0\approx 5$ at $\rho \approx 5\,\rho_0$, the densities characteristic of neutron star cores.

Extrapolations of the chiral FRG equation of state~\cite{DW2015,DW2017} to densities beyond the range realized in neutron stars indicate the existence of a continuous chiral crossover transition around $\rho \gtrsim 8\,\rho_0$. Under such conditions the valence quark cores of the nucleons begin to overlap. It is then an interesting issue how to describe a possible hadron-quark continuity region in an actual model, such as the one developed in Ref.\,\cite{Baym2018}. The sound velocity is again a quantity of prime interest in this context. 

While the present work deals with neutron matter, extensions to symmetric and asymmetric nuclear matter in order to study isospin-dependent Fermi-liquid parameters at high densities are certainly of interest. At densities close to equilibrium nuclear matter, such investigations have previously been performed with nuclear interactions based on chiral effective field theory in conjunction with many-body perturbation theory~\cite{Holt:2011yj}. The nonperturbative chiral FRG approach~\cite{DW2015,DW2017} used in the present work, is in fact designed to be compatible with in-medium ChEFT at low densities, $\rho\lesssim 2\rho_0$, and thus suitable for extrapolations to higher densities. Such considerations might motivate future studies. 

\section*{APPENDICES}
\appendix

\subsection*{A\hspace{3mm}Vector potential contribution to the quasiparticle energy}

Here we refer back to Eq.\,(\ref{eq:qp-int-II}) and present some explicit steps in the derivation of the third term on the right hand side of that equation.  

Consider the variation of the spatial components of the vector field:
\begin{equation}
\frac{\delta \boldsymbol{V}}{\delta n_{p^\prime}}=\frac{\partial \boldsymbol{V}}{\partial\rho}\,\frac{\delta \rho}{\delta n_{p^\prime}}+\frac{\partial \boldsymbol{V}}{\partial\boldsymbol{j}}\,\frac{\delta \boldsymbol{j}}{\delta n_{p^\prime}}~.\nonumber
\end{equation}
The first term becomes:
\begin{equation}
\frac{\partial \boldsymbol{V}}{\partial\rho}=h^\prime(j^2)\,\rho\,\boldsymbol{j}~,\nonumber
\end{equation}
and vanishes in the rest frame where $\boldsymbol{j} = 0$. The derivative in the second term is given by
\begin{equation}
\frac{\partial V_k}{\partial j_m}=h(j^2)\delta_{km}-h^\prime(j^2)\,j_m\,j_k~,\nonumber
\end{equation}
where again the latter part vanishes in the rest frame of the system. Consequently, we find
\begin{equation}\label{eq:variation-V-spatial}
\frac{\delta \boldsymbol{V}}{\delta n_{p^\prime}}=h(j^2)\,\frac{\delta \boldsymbol{j}}{\delta n_{p^\prime}}~.\nonumber
\end{equation}
Using Eqs.\,(\ref{eq:qp-velocity}), (\ref{eq:qp-current}), and (\ref{eq:variation-V-spatial}) the variation of the current yields:
\begin{eqnarray}\label{eq:variation-current-II}
\frac{\delta \boldsymbol{j}}{\delta n_{p^\prime}}&=&\frac{1}{V}\,\boldsymbol{v}_{p^\prime}-\frac{h(j^2)}{V}\sum_p n_p\,\left(\frac{1}{\sqrt{\boldsymbol{p}^2+M^2}}\,\frac{\delta \boldsymbol{j}}{\delta n_{p^\prime}}-\frac{\boldsymbol{p}}{(\sqrt{\boldsymbol{p}^2+M^2})^3}\,\boldsymbol{p}\cdot \frac{\delta \boldsymbol{j}}{\delta n_{p^\prime}}\right)\nonumber\\
&=&\frac{1}{V}\,\boldsymbol{v}_{p^\prime}-\frac{h(j^2)}{V}\sum_p n_p\,\left(\frac{1}{\sqrt{\boldsymbol{p}^2+M^2}}-\frac13\frac{\boldsymbol{p}^2}{(\sqrt{\boldsymbol{p}^2+M^2})^3}\right)\,\frac{\delta \boldsymbol{j}}{\delta n_{p^\prime}}\nonumber\\
&=&\frac{1}{V}\,\boldsymbol{v}_{p^\prime}-\frac{h(j^2)}{V}\sum_p n_p\,\frac{\frac23\boldsymbol{p}^2+M^2}{(\sqrt{\boldsymbol{p}^2+M^2})^3}\,\frac{\delta \boldsymbol{j}}{\delta n_{p^\prime}}\,.\nonumber
\end{eqnarray}
The sum in the last line can be evaluated \cite{Matsui1981}:
\begin{equation}
\frac{1}{V}\sum_p n_p\,\frac{\frac23\boldsymbol{p}^2+M^2}{(\sqrt{\boldsymbol{p}^2+M^2})^3}=\nu\int \frac{d^3p}{(2\pi)^3} \frac{\frac23\boldsymbol{p}^2+M^2}{(\sqrt{\boldsymbol{p}^2+M^2})^3}\,n_p=\frac{\rho}{\sqrt{p_F^2+M^2}}~.\nonumber
\end{equation}
Thus, the variation of the current is given by
\begin{equation}
\frac{\delta \boldsymbol{j}}{\delta n_{p^\prime}}=\frac{\frac{1}{V}\,\boldsymbol{v}_{p^\prime}}{1+\frac{h(j^2)\rho}{\sqrt{p_F^2+M^2}}}=\frac{\frac{1}{V}\,\boldsymbol{v}_{p^\prime}}{1+\frac{V_0}{\sqrt{p_F^2+M^2}}}~.\nonumber
\end{equation}
Finally the contribution of the last term in (\ref{eq:qp-int-II}) to the quasiparticle interaction is
\begin{equation}
f_{pp^\prime}(3)=-h(j^2)\frac{\boldsymbol{p}\cdot\boldsymbol{p}^\prime}{\sqrt{p_F^2+M^2}}\frac{1}{\sqrt{p_F^2+M^2}+V_0}=-h(j^2)\frac{\boldsymbol{p}\cdot\boldsymbol{p}^\prime}{\mu\,\sqrt{p_F^2+M^2}}~,\nonumber
\end{equation}
the result quoted in Eq.\,(\ref{eq:3rdterm}).

\subsection*{B\hspace{3mm}Chiral Nucleon-Meson Field Theory and Functional Renormaliztion Group}

The present study starts from a chiral field theory of mesons and nucleons, based on a linear sigma model with a nonlinear effective potential. The Lagrangian, written in Eq.\,(\ref{eq:ChNM}), is used as UV input for the FRG flow equations at the chiral symmetry breaking scale $\Lambda_\chi \approx 1$ GeV.

The thermodynamics of this model is developed in detail and summarized in Refs.\,\cite{DW2015, DW2017}.
Finite temperatures and chemical potentials are treated using the Matsubara formalism. Minkowski space-time is Wick rotated to Euclidean space. Time components are transformed as $x^0\rightarrow-i\tau$. The $\tau$ dimension is compactified on a circle, such that $\tau$ is restricted to $[0,\beta]$ with the inverse temperature $\beta=1/T$. Time integrals are replaced by $-i\int_0^\beta d\tau$. Boson and fermion fields are periodic or anti-periodic, respectively, under $\tau\rightarrow \tau+\beta$. The Minkowski-space action $S=\int d^4x\;\mathcal L$ is replaced by the Euclidean action $S_{\rm E}=\int_0^\beta d\tau\int d^3x\;{\cal L}_{\rm E}$. 

The FRG scheme proceeds as follows. An effective action, $\Gamma_k[\Phi]$ depending on a renormalization scale $k$, is introduced, where $\Phi$ stands for the set of all chiral boson and nucleon fields. The action $S_{\rm E}=\int_0^\beta d\tau\int d^3x\;{\cal L}_{\rm E} = \Gamma_{k\rightarrow k_{UV}}$  serves as the initialization of $\Gamma_k$ at the UV scale,  $k_{UV} \approx \Lambda_\chi = 4\pi f_\pi$. The flow of $\Gamma_k$ is determined in such a way that it interpolates between the UV action and the full quantum effective action $\Gamma_{\rm eff}=\Gamma_{k=0}$ in the infrared limit, $k\rightarrow 0$. The evolution of $\Gamma_k$ as a function of $k$ is given by Wetterich's flow equation \cite{Wet1993}, schematically written as 
\begin{align}\label{eq:Wetterich}
	\begin{aligned}
		k\,\frac{\partial\Gamma_k[\Phi]}{\partial k}=&
		\begin{aligned}
			\includegraphics[width=0.1\textwidth]{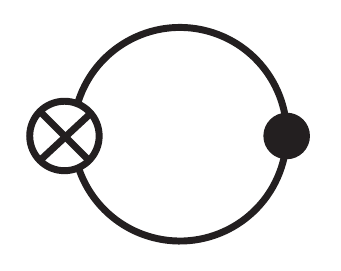}
		\end{aligned} \\ =&\frac 12 {\rm Tr}\left[k\frac{\partial R_k}{\partial k}\cdot\Big(\Gamma_k^{(2)}[\Phi]+R_k\Big)^{-1}\right]~~.
	\end{aligned}
\end{align}
The trace Tr stands for all relevant sums and integrations. A scale regulator, $R_k(p)$, is introduced in order to restrict momenta $p$ in loop integrals to $p^2 < k^2$. The derivative $\partial_k R_k$ has maximum weight at $p^2 \approx k^2$. The matrix $\Gamma_k^{(2)}$ involves second functional derivatives of the effective action with respect to chiral and nucleon fields. It collects the {\it full} inverse propagators of all particles involved. In the pictorial illustration of the flow equation (\ref{eq:Wetterich}) these full propagators are marked by the dot on the loop line while the $k$ regulator is symbolized by the crossed circle.

The welcome feature of the FRG system of equations is its capability to generate fluctuations to all orders beyond mean-field approximation. The ``soft" degrees of freedom that contribute most prominently to these fluctuations  are the pion field with its small mass and low-energy nucleon-hole excitations. Both these types of excitations enter non-perturbatively through their full propagators in the flow equation (\ref{eq:Wetterich}). 

For the treatment of a dense and thermal medium with inclusion of fluctuations it is useful to compute the flow of the difference between the effective action at given values of temperature and chemical potential, $\Gamma_k(T,\mu)$, as compared to the potential at a reference point for which we choose equilibrium nuclear matter at zero temperature, $\Gamma_k(0,\mu_0)$ with $\mu_0 = M_N + E_0/A = 923$ MeV. The flow of the difference, $\bar\Gamma_k=\Gamma_k(T,\mu)-\Gamma_k(0,\mu_0)$, satisfies the FRG equation
\begin{align}
	\begin{aligned}
		\frac{k\,\partial\bar \Gamma_k}{\partial k}(T,\mu)&=
		\begin{aligned}
			\hspace{-.1cm}
			\vspace{1cm}
			\includegraphics[width=0.1\textwidth]{wetterich_fermion}
		\end{aligned} \vspace{-1cm}\Bigg|_{T,\mu}-
		\begin{aligned}
			\hspace{-.1cm}
			\vspace{1cm}
			\includegraphics[width=0.1\textwidth]{wetterich_fermion}
		\end{aligned} \Bigg|_{\begin{subarray}{l} T=0 \\ \mu=\mu_0 \end{subarray}}.
	\end{aligned}
\end{align}
The actual computational work involves some simplifying assumptions and approximations: the effective action is treated in leading order of the derivative expansion and we work in the local potential approximation, neglecting (small) wave function renormalization effects on the chiral boson fields  and possible higher order derivative couplings. Moreover, the $k$ running of the Yukawa coupling $g$ is ignored; the dependence of the nucleon mass on temperature and chemical potential scales with that of the in-medium pion decay constant, $f_\pi^*(T,\mu) = \langle\sigma\rangle(T,\mu)$. 

The nucleon mass $M = g\sigma$ is coupled dynamically to the scalar field. Its expectation value, $\langle\sigma\rangle$ is normalized to the pion decay constant $f_\pi$ in the vacuum so that the mass of the free nucleon satisfies the Goldberger-Treiman relation, $M_0 = g\,f_\pi$. The coupling $g\approx 10$ stands for the ratio of pion-nucleon coupling constant, $g_{\pi N}$, and the axial vector coupling constant, $g_A$, of the nucleon: $g = g_{\pi N}/g_A$. In general, $\langle\sigma\rangle$ acts as an order parameter for spontaneously broken chiral symmetry. The region of  temperatures $T$ and densities $\rho$ where this chiral FRG framework can be applied is defined by nonzero $\langle\sigma\rangle(T,\rho)\equiv f^*_\pi(T,\rho)$. The full FRG calculations \cite{DW2015,DW2017} point out that the domain with $\langle\sigma\rangle\neq 0$ covers the broad range $0<T\lesssim 100$ MeV and $0<\rho\lesssim 7\,\rho_0$. In contrast, using mean-field approximation would generate a first-order chiral phase transition around $\rho\approx 3\,\rho_0$.

\subsection*{C\hspace{3mm}Chiral FRG equation of state and quasiparticle properties: numerical details}

\subsubsection{Pressure}
For practical purposes and applications, the chiral EoS of neutron star matter, $P(\cal E)$ shown in Fig.\,\ref{fig:1}, is given here in a parametrized form of an accurate Pad\'e fit in the baryon density range $0.1\leq \rho/\rho_0 \leq 6$ (with $\rho_0 = 0.16$ fm$^{-3}$): 
\begin{equation}
\frac{P(z)}{ \mathrm{MeV/fm^3}} = \frac{\sum_{n=3...5} a_n z^n}{ 1+\sum_{m=1}^5 b_m z^m}~,~~~~~~z = \frac{{\cal E}}{ \mathrm{GeV/fm^3}}~,
\end{equation}
 with the parameters
\begin{eqnarray}
a_3 &=& 1270.28~~~~~a_4=7236.08~~~~~a_5=5.99\nonumber \\
b_1=1.7857~~~~~b_2&=&26.4331~~~~~b_3=-3.0629~~~~~b_4=11.8366~~~~~b_5=-3.9431~.
\end{eqnarray} 

\subsubsection{Energy density}
A further useful quantity to work with is the energy density ${\cal E}(\rho)$ as a function of baryon density, also presented in terms of a Pad\'e fit for practical computations:

\begin{equation}
\frac{{\cal E}(y)}{\mathrm{GeV/fm^3}}=\frac{c_1 y+c_2 y^2}{ 1+ \sum_{n=1}^4 d_n y^n}~,~~~~~y=\rho/ \rho_0~, ~~~~~(\rho_0 = 0.16\,{\mathrm fm^{-3}})~,
\end{equation}
with the coefficients:
\begin{eqnarray}
c_1&=& 0.1509~~~~~~~c_2=0.0711\nonumber\\
d_1=0.4857~~~~~d_2&=&-0.0343~~~~~d_3=0.0020~~~~~d_4 = 6.5\times 10^{-5}~.
\end{eqnarray}

\subsubsection{Quasiparticle mass}

The density-dependent baryon mass $M(\rho)$ is a key ingredient of the quasiparticle energy and chemical potential,
\begin{equation}
\varepsilon_{p=p_F} = \mu = \sqrt{p_F^2 + M^2(\rho)} + U(\rho)~.\nonumber
\end{equation}

An excellent numerical fit to $M(\rho)$, resulting from the chiral FRG calculation of \cite{DW2015,DW2017}, is found using a Pad\'e representation as follows:
\begin{equation}
\frac{M(y)}{ M_0} = \frac{\langle\sigma\rangle}{ f_\pi} = \frac{1+\sum_{n=1}^3\alpha_n\,y^n}{ 1+\sum_{n=1}^3\beta_n\,y^n}~,~~~~~y = \frac{\rho}{\rho_0}~,\nonumber
\end{equation}
with
\begin{eqnarray}
\alpha_1 &=& 0.6170~,~~~\alpha_2 = -0.1462~,~~~\alpha_3=0.0052~,~~~\nonumber\\
\beta_1&=& 0.8699~,~~~\beta_2 = -0.0821~,~~~\beta_3 = -0.0076~.\nonumber
\end{eqnarray}

Table \ref{t1} collects numerical values of $P(\cal E)$ together with selected quasiparticle properties for baryon densities in the range $(0.08 - 0.96)$fm$^{-3}$.  
\begin{table}[tbh]
\caption{Chiral FRG equation of state \cite{DW2015, DW2017} for neutron star matter at $T = 0$: baryon density $\rho$ in units of $\rho_0 = 0.16$ fm$^{-3}$, energy density  ${\cal E}$, pressure $P$, baryon chemical potential $\mu = \frac{\partial {\cal E}}{ \partial\rho}$, density-dependent quasiparticle mass $M(\rho)$ [with $M_0 = M(\rho = 0) = 939.6$ MeV], and Landau effective mass $m^*(\rho) = \sqrt{p_F^2 + M^2(\rho)}$. }
\label{t1}
\begin{center}
\begin{tabular}{ c c c c c c }
\hline
$~~\rho/\rho_0~~~~$ &${\cal E}$\,[GeV/fm$^{3}$]~~~~&$P$\,[MeV/fm$^{3}$]~~~~~&$\mu(\rho)$\,[GeV]~~~& $M(\rho)$\,[MeV]~~~&$m^*(\rho)$\,[MeV]\\
\hline
 0.5 & 0.076 & 0.62 & 0.9509 & 845.9 & 885.9 \\ 
 1.0 & 0.153 & 4.51 & 0.9829 & 779.1  & 846.6\\  
 1.5 & 0.233 & 13.15 & 1.0259 & 724.2 & 817.5\\
 2.0 & 0.317 & 26.84 & 1.0746 & 676.0 & 794.6 \\ 
 2.5 & 0.405 & 45.53 & 1.1266 & 632.1  & 775.8\\
 3.0 & 0.497 & 69.14 & 1.1805 & 591.0 & 760.1\\
 3.5 & 0.594 & 97.89 & 1.2356 & 551.7  & 746.7\\
 4.0 & 0.695 & 131.39& 1.2914  & 513.3  & 735.0\\
 4.5 & 0.801 & 169.89 & 1.3477 & 474.6 & 724.3\\
 5.0 & 0.911 & 212.99 & 1.4044 & 433.9 & 713.7\\
 5.5 & 1.025 & 260.73& 1.4616 & 388.0 & 702.0\\
 6.0 & 1.145 & 314.23 & 1.5195  & 329.0 & 686.2\\
\hline   
\end{tabular}
\end{center}
\end{table}

{\bf ACKNOWLEDGEMENTS}
\vspace{0.5cm}

One of the authors (W.W.) gratefully acknowledges fruitful periods as visiting professor at the ExtreMe Matter Institute (EMMI) of GSI-FAIR in Darmstadt and at the Physics Department of the University of Tokyo. He thanks Kenji Fukushima for his kind hospitality and many stimulating discussions. Helpful comments and suggestions by Norbert Kaiser are also much appreciated.

\end{document}